\documentclass[prl, twocolumn, amsmath,superscriptaddress]{revtex4}
\usepackage{dcolumn}
\usepackage{bm}
\usepackage{graphicx}
\usepackage{color}
\usepackage{xcolor}
\usepackage{amsmath}
\usepackage{amssymb}
\usepackage{dcolumn}
\usepackage{epsfig}
\usepackage{hyperref} 
\DeclareGraphicsExtensions{. jpg,. pdf, . mps, . png, . eps, . ps, . EPS}

\newcommand{\be}{\begin{equation}}
\newcommand{\ee}{\end{equation}}

\definecolor{myred}{RGB}{168,5,14}
\definecolor{myblue}{RGB}{13,13,255}
\definecolor{editorcolor}{RGB}{200,5,14}
\definecolor{mygreen}{RGB}{20,150,20}

\begin{document}

\title{Topologically protected synchronization in networks}

\author{Massimo Ostilli}
\affiliation{
Instituto de
    F\'isica, Universidade Federal da Bahia, Salvador,    Brazil}

\begin{abstract}
  In a graph, we say that two nodes are topologically equivalent if their sets of first neighbors, excluding the two nodes, coincide.
  We prove that nonlinearly coupled {heterogeneous} oscillators located on a group of topologically equivalent nodes can get easily synchronized when
    the group forms a fully connected subgraph (or combinations {thereof}), regardless of the status of all the other oscillators.
  More generally, any change occurring in the remainder of the graph will not alter the synchronization status of the group.
  Typically, the group can synchronize when $k^{(\mathrm{OUT})}\leq k^{(\mathrm{IN})}$, $k^{(\mathrm{IN})}$ and $k^{(\mathrm{OUT})}$ being the common internal and outgoing
  degree of each node in the group, respectively. Simulations confirm our rigorous analysis and suggest that groups of topologically equivalent nodes
  {act as} independent pacemakers.
  \end{abstract}

\maketitle

Many physical, biological, chemical, and technological systems
can effectively be seen as networks of interacting oscillators, each with its own natural frequency.
Spontaneous synchronization represents one of the most intriguing and ubiquitous aspects of these systems~\cite{StrogatzC,Pikovsky,ArenasReview} and has countless applications
such as, pacemaker cells in the heart~\cite{Heart}, pacemaker cells in the brain~\cite{Brain,Brain2},
ﬂashing ﬁreﬂies~\cite{Fireflies}, arrays of lasers~\cite{Lasers}, and superconducting Josephson junctions~\cite{Josephson}.
Put simply, if the coupling constant $J$ is larger than
the spread $\Delta \omega$ of the system's natural frequencies, a finite portion of its oscillators synchronizes, \textit{i.e.},
they rotate according to a common frequency and their phases fall into step with one another. 
{The Kuramoto model (KM)~\cite{Kuramoto1,Strogatz2000}
it is to the theory of synchronization as the Ising model is to the theory of phase transitions.}
In its mean-field version, 
the KM allows to be exactly solved~\cite{Strogatz2000} while, in general, despite no exact solution is known,
rigorous bounds~\cite{Strogatz1988},
simulations, and theoretical studies~\cite{ArenasReview,Munoz,Kahng}
have
succeeded to provide a general description of the different synchronization scenarios taking place as a function
of the topology of the underlying graph~\cite{Goltsev}, ranging from regular to random, small-world, and scale-free.
For the present manuscript, we emphasize the role played by modularity.
Many complex networks are modular, \textit{i.e.}, can be {partitioned into subgraphs, also known as communities,
with different internal
and external connectivities~\cite{Barabasi_Hierarchical,Santo}.
It has been understood that, if $J>\Delta \omega$, the densely connected communities} synchronize first and, subsequently, the larger and less densely connected ones
also tend to synchronize, until full synchronization is achieved~\cite{ArenasPRL,ArenasReview}. 

The above literature analyzes synchronization
in the thermodynamic limit as a phase transition, with
a phase where each oscillator tends to rotate with its own natural frequency, and a phase where most of the oscillators
rotate with a common frequency.

Here, we consider a rather different issue related to a local and protected synchronization: ``local'' because it concerns a finite group of oscillators embedded in some arbitrary
graph containing the group;
``protected'' because the remainder of the graph cannot affect the synchronization status of the group, even under noise.
{Given a graph with $N$ nodes, consider a group of $N'\leq N$ nodes and split the set of neighbors of each of these $N'$ nodes into two sets:
  those contained within the group and those external to the group (external sets).
We say that the $N'$ nodes are topologically equivalent (TE) if they have the same number of links and identical external sets.
In this work, we prove that, in the KM, a group of TE nodes can synchronize when
the group forms a fully connected (FC) subgraph, regardless of the status of the other oscillators. We provide the exact conditions
under which this occurs.
  Typically, the group can synchronize when $k^{(\mathrm{OUT})}\leq k^{(\mathrm{IN})}$, where $k^{(\mathrm{IN})}$ and $k^{(\mathrm{OUT})}$ are the common internal and outgoing
  degrees of each node in the group.} 

{Formally, this work falls within synchronization of chaotic systems via bidirectional couplings~\cite{Boccaletti}, a less-developed branch compared to the unidirectional case,
  where a common driver acting on uncoupled oscillators can induce synchronization~\cite{Teramae,Feng,Hata,Boccaletti2}.
  In the bidirectional case, analytical studies mainly concern mean-field models\cite{Strogatz2000,Rosenblum1,Rosenblum2}; however,
  Pecora \textit{et al.}~\cite{Pecora} showed that clusters of identical coupled oscillators can synchronize when they exhibit specific graph symmetries.
  On the other hand, MacArthur \textit{et al.}~\cite{MacArthur} demonstrated that such symmetries
  --- contrary to simple models of graph theory, like the random graph or the Barab\'asi-Albert model ---
  are very common in real-world networks. Moreover, \cite{MacArthur} showed that most of these symmetric clusters are symmetric cliques, matching our definition of FC TE nodes.
  {More recently, the transition toward complete synchronization through intermediate synchronized clusters
    have received further attention and the general mechanism has become better understood, especially
    in models involving identical oscillators \cite{Lin,Lambiotte,Fiore,Gambuzza,Multilayer,Boccaletti3}.

The novelty of our work lies in the derivation of rigorous bounds for the phase differences of a group (or cluster) of coupled oscillators governed by a generic Kuramoto model with heterogeneous frequencies. These bounds depend solely on the model parameters and are independent of the dynamical states of the other oscillators.    
  Unlike the works cited above, where the identity of the oscillators is a fundamental assumption for the analytical treatment,
  our KM-based analysis allows heterogeneous frequencies and supports multiple coexisting synchronized
manifolds (more precisely, nearly-synchronized manifolds~\cite{Note2}), as observed in more realistic networks, particularly in biological systems.}

{We derive  our result starting from the simplest group and then generalize.}
Given a graph with $N$ nodes and symmetric adjacency matrix {$a_{i,j}$},
consider a KM on the top of it with coupling $J>0$: the phases of $N$ oscillators, $\theta_1,\ldots,\theta_N$, with natural frequencies $\omega_1,\ldots,\omega_N$, evolve
according to 
\begin{eqnarray}
\label{Kuramoto}
\dot{\theta}_i=\omega_i+J\sum_{j}a_{i,j}\sin\left(\theta_j-\theta_i\right).
\end{eqnarray}

\textit{Group of $N'=2$ TE nodes.}
Consider now two specific nodes, say node 1 and node 2 (note that $a_{1,2}=a_{2,1}$).
From now on, unless otherwise stated, we assume that the two nodes are connected: $a_{1,2}=1$.
For the phase difference variable ${\Theta}=\theta_2-\theta_1$, Eqs. (\ref{Kuramoto}) lead to
\begin{eqnarray}
\label{Kuramoto4}
&& \dot{{\Theta}}=\omega_2-\omega_1-2J\sin\left({\Theta}\right)+J\left[ h_2(\theta_2)- h_1(\theta_1)\right],
\end{eqnarray}
where we have introduced the two functions{\cite{Note3}}
\begin{eqnarray}
\label{Kuramoto5a}
&& h_i(\theta)=\sum_{j\neq 1,2}a_{i,j}\sin\left(\theta_j-\theta\right), \quad i=1,2.
\end{eqnarray}
Let us now suppose that node 1 and node 2 are TE, \textit{i.e.}, they ``see'' the
same remaining graph: 
\begin{eqnarray}
\label{Kuramoto2}
a_{1,j}=a_{2,j}, \quad \forall j\neq 1,2. 
\end{eqnarray}
In this case 
\begin{eqnarray}
\label{Kurah}
h(\theta)=h_1(\theta)=h_2(\theta),
\end{eqnarray}
and Eq. (\ref{Kuramoto4}) becomes
\begin{eqnarray}
\label{Kuramoto5}
&& \dot{{\Theta}}=\omega_2-\omega_1-2J\sin\left({\Theta}\right)+J\left[ h(\theta_2)- h(\theta_1)\right].
\end{eqnarray}
We can rewrite Eq. (\ref{Kuramoto5}) as
\begin{eqnarray}
\label{Kuramoto5b}
&& \dot{{\Theta}}=\omega_2-\omega_1-2J\sin\left({\Theta}\right)+J h'({\theta}^*){\Theta}
\end{eqnarray}
where ${\theta}^*\in(\min\{\theta_1,\theta_2\},\max\{\theta_1,\theta_2\})$ and we have applied the mean-value theorem to the function $h(\theta)$.
Note that ${\theta}^*$ is itself an unknown function of $\theta_1$ and ${\Theta}$ (or, alternatively, $\theta_2$ and ${\Theta}$), yet, as we shall see in a moment, it is worth to consider Eq. (\ref{Kuramoto5b}).
We will use the bounds:
\begin{eqnarray}
\label{boundout}
&& {-k^{(\mathrm{OUT})}\leq } h'(\theta)\leq k^{(\mathrm{OUT})}, \quad \forall \theta,
\end{eqnarray}
where $k^{(\mathrm{OUT})} $ is the (common) outgoing degree of the two nodes
\begin{eqnarray}
\label{kout}
k^{(\mathrm{OUT})}=\sum_{j\neq 1,2} a_{1,j}=\sum_{j\neq 1,2} a_{2,j},
\end{eqnarray}
and
\begin{eqnarray}
\label{boundin}
-\theta< -\sin(\theta)<-\theta(1-\gamma), \quad 0 < \theta<1,
\end{eqnarray}
where {\cite{SM}}
\begin{eqnarray}
\label{gammadef}
\gamma=\frac{1}{3!}+\frac{1}{7!}+\frac{1}{11!}+\ldots=0.1668651044\ldots.
\end{eqnarray}

\textit{Equal frequencies with $a_{1,2}=1$.}
Let us assume that the two nodes have equal frequencies, $\omega_1=\omega_2$.
From Eq. (\ref{Kuramoto5b}) we see that the manifold $\theta_1=\theta_2$ ($=\theta^*$), \textit{i.e.}, 
${\Theta}=~0$, provides a fixed point.
Let us introduce $\lambda=k^{(\mathrm{OUT})}-2(1-\gamma)$ and $\mu=k^{(\mathrm{OUT})}+2$.
From Eq. (\ref{Kuramoto5b}), by using the bounds (\ref{boundout}) and (\ref{boundin}), we see that, if $1>{\Theta}> 0$,
\begin{eqnarray}
\label{Rigor1}
&& -J \mu{\Theta} < \dot{{\Theta}} < J \lambda {\Theta}, 
\end{eqnarray}
while, if $-1<{\Theta}<0$, hold the opposite inequalities.
Let us assume that, for the initial condition $\Theta_0=\Theta(t=0)$, we have $0<\Theta_0<1$.
By continuity, there exists a sufficiently small time $t_1$
such that $0<{\Theta}(t)<1$ for $t\in[0,t_1)$ whereby, from Eq. (\ref{Rigor1}), for $t\in[0,t_1)$ we get
\begin{eqnarray}
\label{Rigor2}
&& {\Theta}_0 e^{-J \mu t} < {\Theta} < {\Theta}_0 e^{J \lambda t}. 
\end{eqnarray}
On the other hand, if $\lambda<0$, Eq. (\ref{Rigor2}), and the fact that $\theta_i(t)$'s and
$\dot{\theta}_i(t)$'s are continuous, imply that $t_1=\infty$. 
{One can check this by a graphical construction, but also analytically~\cite{SM}}.
For simplicity, in the subsequent cases, on assuming certain constrains on the initial conditions,
we shall limit ourselves to check that the upper bounding solution satisfies the same constrains for any $t$. Note however that, in general,
simply bounding $|\Theta|$ is not enough for claiming good synchronization;
the conservation of the sign of $\Theta$
is crucial.

In conclusion, within a basin of attraction for the initial conditions
$|{\Theta}_0|<1$,
a sufficient condition for the fixed point ${\Theta}=0$ to be stable is $k^{(\mathrm{OUT})}<2(1-\gamma)$.
Of course, since $1<2(1-\gamma)<2$ and $k^{(\mathrm{OUT})}$ is integer, it follows that $k^{(\mathrm{OUT})}$ can be either 0 or 1,
but, in view of generalizations to subsystems with $N'>2$ TE oscillators,
it is useful to retain the general inequality.
Note that this synchronization between node 1 and 2 occurs regardless of the dynamics of all the other oscillators, which in particular do not need to be synchronized
and can have arbitrary frequencies.
We stress that this is a consequence of the topological equivalence (\ref{Kuramoto2}); without it, $h_1$ and $h_2$ 
do not cancel out in the manifold $\theta_1=\theta_2$ in Eq. (\ref{Kuramoto4}).
We can even imaging to modify the inner part of the remainder of the graph by dynamically removing,
adding, or rewiring some of its links, as well as by allowing for the presence of any site-dependent noise:
as far as such links are not those arriving at nodes 1 and 2, and as far as such noise applies only to the other nodes,
the sufficient condition $k^{(\mathrm{OUT})}\leq 2(1-\gamma)$ remains satisfied, \textit{i.e.},
the subsystem fixed point $\theta_1=\theta_2$ keeps being stable. 

\textit{Equal frequencies with $a_{1,2}=0$.}
Before analyzing the most general case, it is worth also considering the sub-case in which the two nodes are disconnected,\textit{i.e.}, $a_{1,2}=0$ and have equal frequencies.
In this case, Eq. (\ref{Kuramoto5b}) is of no help because $\theta_1=\theta_2$ is no longer a stable attractor,
in other words, in general, the two oscillators do not get synchronized.
However, from Eq. (\ref{Kuramoto})
we see that, if nodes 1 and 2 are TE, $\theta_1$ and $\theta_2$ obey the same identical equation: we simply have that one oscillator follows the other along the same
trajectory and, in particular, their phases will remain equal at any instant in the case of same initial conditions.
In this specific case, ${\Theta}=0$ is no longer an attractor but rather a constant of the motion.

\textit{Different frequencies with $a_{1,2}=1$.}
Let us now allow for the two frequencies to be different.
If the pair is isolated, \textit{i.e.}, $k^{(\mathrm{OUT})}=0$, we have $h'(\theta^*)=0$ and
Eq. (\ref{Kuramoto5b}) returns the known critical condition for the synchronization of ${N=}N'=2$ oscillators: $2J>|\omega_2-\omega_1|$, the stability condition 
of the fixed point ${\Theta}$ being $\cos({\Theta})>0$.
In the general case, $h'(\theta^*)\neq 0$, 
although Eq. (\ref{Kuramoto5b}) admits a formal fixed point, the fact that $\theta^*$ depends on $\theta_1$ (besides ${\Theta}$),
which in turn is a function of time, makes this formal fixed point useless. Our approach here is different:
we assume that both {$|\omega_2-\omega_1|/J$} and the initial condition $|{\Theta}_0|=|\theta_2(t=0)-\theta_1(t=0)|$ are sufficiently small and seek a bounding solution
by exploiting again the bounds (\ref{boundout}) and (\ref{boundin}).
Let us suppose ${\Omega}=\omega_2-\omega_1>0$ and $0<{\Theta}_0<1$.
By continuity, at small enough times, we also have $0<{\Theta}<1$ and
from Eq. (\ref{Kuramoto5b}) we get 
\begin{eqnarray}
\label{bound2}
&& \dot{{\Theta}} \leq {\Omega} + J\lambda{\Theta}. 
\end{eqnarray}
Eq. (\ref{bound2}) implies that, if $\lambda=k^{(\mathrm{OUT})}-2(1-\gamma)<0$, ${\Theta}$, with initial condition $0<{\Theta}_0<1$, remains bounded as
\begin{eqnarray}
\label{bound3}
&& {\Theta} \leq \frac{{\Omega}}{J|\lambda|} + \left[{\Theta}_0-\frac{{\Omega}}{J|\lambda|}\right]\exp\left[-J|\lambda|t\right].
\end{eqnarray}

Note that, for $0<\Omega/J|\lambda|<\Theta_0$, all the above procedure turns out to be consistent with the required bound $0<{\Theta}<1$ for any $t$.

In conclusion, we have proven that, under the three conditions $k^{(\mathrm{OUT})}<2(1-\gamma)$ (equivalent to $k^{(\mathrm{OUT})}\leq 1$),
$0<{\Theta}_0<1$, and
$0<\Omega/J|\lambda|<\Theta_0$,
we have
$0\leq \Theta\leq \Omega/J|\lambda|$
when $t\to\infty$.
In other words, the phases of the two oscillators get asymptotically
close to each other and the larger is $J$ the closer they stay.
Moreover, as in the case of equal frequencies,
we see that this bound holds regardless of the status of all the other oscillators
located on the remainder of the graph and, again, any change in it, cannot affect the synchronization of the pair.

It is important to note that the nature of the condition $k^{(\mathrm{OUT})}\leq 1$, unlike the others related to the initial value of ${\Theta}$ and to the ratio $|\Omega|/J$,
is strictly topological. We shall call it the topological condition (TC).

\textit{Generalization to $N'>2$ TE oscillators.}
Let us now consider a subsystem of $N'=3$ TE nodes, say nodes 1, 2 and 3. Besides being TE with respect to the remainder of the graph, we want them
to be TE among each other;
in particular, each of them must have the same number of links pointing to the other two TE nodes. For $N'=3$ there exists only one possibility,
the one where $k^{(\mathrm{IN})}=2$, \textit{i.e.}, the three nodes form a triangle.
{As anticipated in the Introduction}, 
$k^{(\mathrm{IN})}$ is the (common) number of links emanating from each node of the group and pointing to the other nodes of the same group.
Note that, formally, both the subsystem $N'=3$ and the already seen case $N'=2$ (where it was $k^{(\mathrm{IN})}=1$),
are FC graphs with $N'$ nodes.
However, as Fig. \ref{figIllustrative} shows, when $N'>3$, there exist more configurations in which the $N'$ TE nodes can be arranged. In fact, the number of possible arrangements
tends to grow exponentially with $N'$, but with a smaller rate for $N'$ odd (see also \cite{SM}).

\begin{figure}[htb]
  \centering
  \includegraphics[width=0.45\columnwidth,clip]{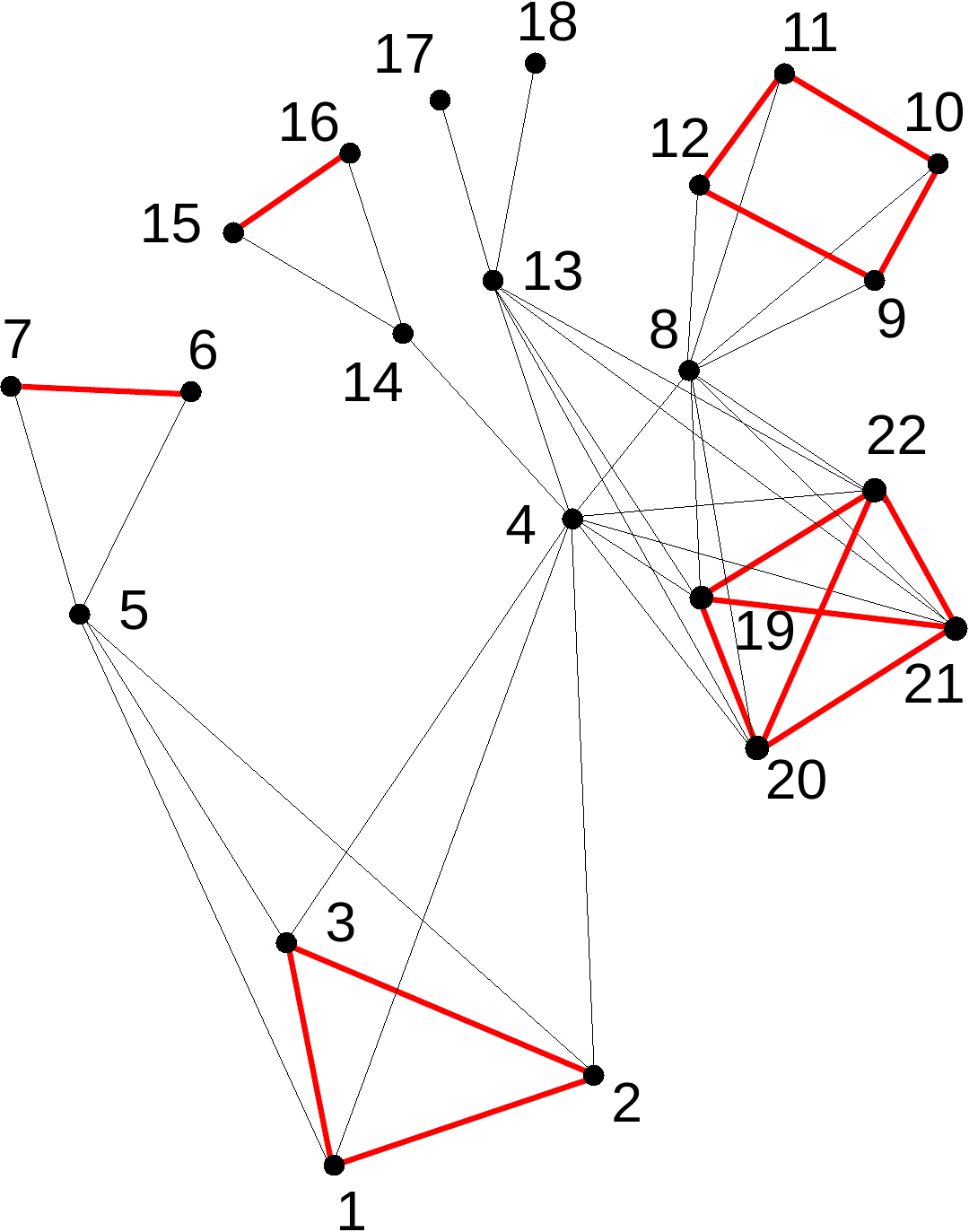}
  \caption{Graph associated to model (\ref{Kuramoto}) with $N=22$
    nodes and {$L=46$ links (15 red-ticker and 31 black-thin)}. The graph contains five groups of TE nodes:
    one FC with $N'=3$ (nodes 1, 2 and 3), two FC with $N'=2$
    (nodes 7 and 6, and nodes 15 and 16, respectively);
    one with $N'=4$ forming a square (nodes 9, 10, 11, and 12); and one FC with $N'=4$ (nodes 19, 20, 21, 22).
    In each group, the internal links are drawn as ticker (red).
    In the four FC groups we have $k^{(\mathrm{OUT})}\leq k^{(\mathrm{IN})}$ while in the other group
    we have $k^{(\mathrm{OUT})}< k^{(\mathrm{IN})}$, hence, in each group, the TC for synchronization are satisfied.
    }
  \label{figIllustrative}
\end{figure}  

We introduce some notation and point out a few crucial points.
Given $N'$ TE nodes with indices $1,2,\ldots,N'$, we indicate their corresponding phase differences by ${\Theta}_{i,j}=\theta_i-\theta_j$
and frequency differences by ${\Omega}_{i,j}=\omega_i-\omega_j$.
Observe that ${\Theta}_{j,i}=-{\Theta}_{i,j}$ and that the variables $\{{\Theta}_{i,j}\}$ with $j>i$, are not all independent.
For example, for $N'=3$ we have the constrain
${\Theta}_{3,1}={\Theta}_{3,2}+{\Theta}_{{2,1}}$.
In general, given $N'$, we can always write a system of $N'-1$ independent equations involving $N'-1$ independent variables.
We shall also assume $0<{\Theta}_{i,j}(0)<1/(N'-1)$ so that, at small enough times, we also have $0<{\Theta}_{i,j}(t)<1/(N'-1)$. Our general strategy is to
use the bounds (\ref{boundout}), (\ref{boundin}); 
if the found bounding solution satisfies the above constrains for any $t$, the procedure is consistent.
The resulting bounding system can be written vectorially:
\begin{align}
  \label{bbb}
\dot{\bm{{\Theta}}}\leq \bm{{\Omega}} + J \bm{B}\cdot \bm{{\Theta}},
\end{align}
where $\bm{{\Theta}}=({\Theta}_{2,1},{\Theta}_{3,2},\ldots,{\Theta}_{N',N'-1})^T$, $\bm{{\Omega}}=({\Omega}_{2,1},{\Omega}_{3,2},\ldots,{\Omega}_{N',N'-1})^T$,
and $\bm{B}$ is a $(N'-1)\times (N'-1)$ matrix that depends on the parameters
$N'$, $k^{(\mathrm{OUT})}$, $k^{(\mathrm{IN})}$, and $\gamma$.
Let $\lambda_i$ and $\bm{u}_i$ be the eigenvalues and normalized eigenvectors of $\bm{B}$, respectively. The coefficients $c_i(t)=\bm{{\Theta}}\cdot \bm{u}_i$
satisfy the decoupled system $d c_i(t)/dt \leq \bm{{\Omega}}\cdot \bm{u}_i+J\lambda_i c_i(t)$. As a consequence, each component of the vector $\bm{{\Theta}}$ remains bounded
if all the eigenvalues of $\bm{B}$ are negative, leading to the TC
(note that the {$\lambda$'s} do not depend on $J$ or $\bm{{\Omega}}$).
For example, in the FC case $N'=3$ we have
\begin{align}
  \bm{B}=
  \begin{bmatrix}
    k^{(\mathrm{OUT})}-3(1-\gamma)       & \gamma \\
    \gamma              & k^{(\mathrm{OUT})}-3(1-\gamma),
  \end{bmatrix}
\end{align}
and the eigenvalues of $B$ are
$\lambda_1=k^{(\mathrm{OUT})}-3 +4\gamma$ 
and $\lambda_2=k^{(\mathrm{OUT})}-3 +2\gamma$. 
We conclude that the group of $N'=3$ TE FC nodes synchronize if $k^{(\mathrm{OUT})}-3 +4\gamma<0$, which, on observing that
$0<4\gamma<1$, amounts
to $k^{(\mathrm{OUT})}\leq 2$, or $k^{(\mathrm{OUT})}\leq k^{(\mathrm{IN})}$, where we have made use of the fact that, here, $k^{(\mathrm{IN})}=2$.
Note that also for the previously seen case $N'=2$, where it was $k^{(\mathrm{IN})}=1$, the TC for the synchronization can be written as
$k^{(\mathrm{OUT})}\leq k^{(\mathrm{IN})}$. In general, for a group of $N'$ FC TE nodes, the diagonal elements of the matrix $\bm{B}$ are all equal to
$k^{(\mathrm{OUT})}-N'(1-\gamma)$, while the off-diagonal ones are in the form $p\gamma$ where $p\in \{1,\ldots,N'-2\}$. 
We have verified that for $N'=4$ the TC remains $k^{(\mathrm{OUT})}\leq k^{(\mathrm{IN})}$ but for $N'=5$ the TC becomes $k^{(\mathrm{OUT})}\leq k^{(\mathrm{IN})}-1$.
For details see \cite{SM} where we also analyze cases where the group forms regular polygons. 
We stress that these are sufficient conditions; they might be not necessary.
\begin{figure}[htb]
  \centering
  \includegraphics[width=0.92\columnwidth,clip]{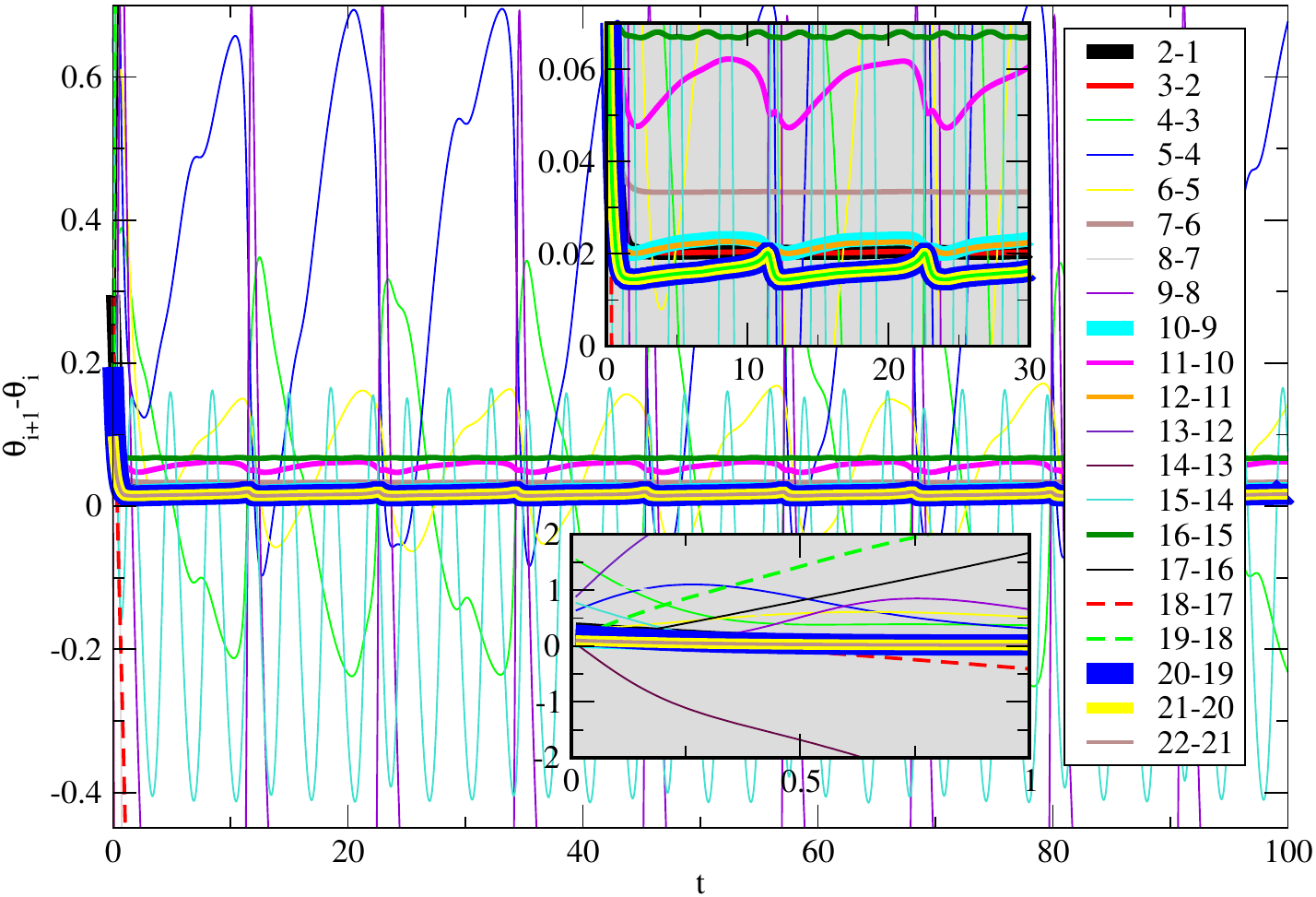}
 \caption{Numerical solution of model (\ref{Kuramoto}) with $J=1$ and $N=22$ oscillators on the top of the graph of Fig. \ref{figIllustrative}.
   Plots correspond to the variables  $\theta_{i+1}-\theta_i$, for $i=1,\ldots,N-1$.
   In each group of TE nodes, the initial conditions, as well as the natural frequencies, are chosen very close to each other as to guarantee
   the strict sufficient conditions explained in the text. 
   It is evident that for several indices $i$, $\theta_{i+1}-\theta_i$ drifts away, while
   there exists a group of indices where $\theta_{i+1}-\theta_i$ remains bounded. As can be better checked by the Insets, 
   the bounded variables includes all the five groups of TE nodes and a few others which, however, are manifestly less synchronized than the TE nodes.
   Note in particular that, unlike the latter, in each group, as analytically predicted, the sign of $\theta_{i+1}-\theta_i$ does not change over time.
   {See Fig. \ref{fig2_SM} and \cite{SM} for further visualization}.
    }
  \label{figIllustrative2}
\end{figure}

We illustrate our analysis by solving numerically~\cite{Numerical_Recipes} model (\ref{Kuramoto}) in a system of moderate size but involving most of the ideas so far discussed.
The graph associated to this system is depicted in Fig. \ref{figIllustrative}.
It contains five groups of TE nodes of various kind for which the TC is satisfied.
We performed several simulations with different choices of the parameters $\omega_1,\ldots,\omega_N$ and $\theta_1(0),\ldots,\theta_N(0)$ ranging from situations
in which the sufficient conditions for synchronization discussed previously are strictly satisfied, to situations in which only the TC is satisfied.
Remarkably, even in these latter situations, the synchronization scenario predicted analytically (via stricter conditions) holds.
Figs. \ref{figIllustrative2} and \ref{figIllustrative3} show the behavior of $\theta_{i+1}(t)-\theta_i(t)$ in the two situations.
Inset of Fig. \ref{figIllustrative3} shows also the rotating numbers $\theta_i(t)/t$~\cite{StrogatzC}. For more details see also Fig. \ref{fig2_SM} and \cite{SM}.                         
\begin{figure}[htb]
  \centering
  \includegraphics[width=0.92\columnwidth,clip]{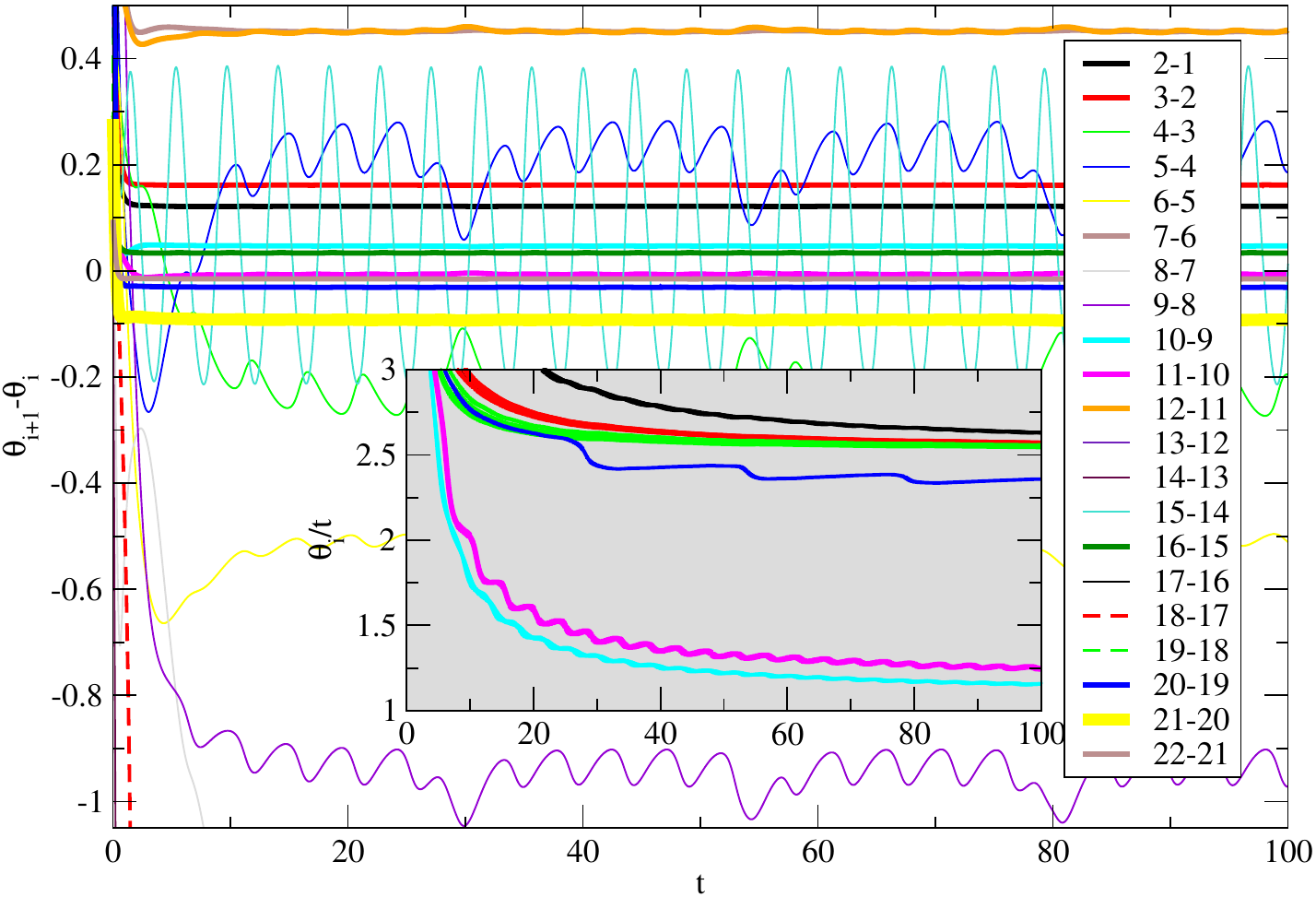}
  \caption{As in Fig. \ref{figIllustrative2} but without strict conditions.  
       Note that $\theta_{i+1}-\theta_i$ becomes almost constant and with constant sign whenever $i$ belongs to a group of TE nodes (unlike the case with strict
       conditions, a change of sign can occur at short times as for the plot of $\theta_{21}-\theta_{20}$).
       Inset: rotating numbers for the same system. Asymptotically, there emerge six sets of converging lines. Each set
       is indicated by different colors as follows (top to bottom): black (nodes 1, 13, 19, 20, 21, 22); red (nodes 2, 3, 4, 5, 6, 7); green (nodes 8, 9, 10, 11, 12);
       blue (node 17); magenta (node 18); cyan (nodes 14, 15, 16).
    }
  \label{figIllustrative3}
\end{figure}

      \begin{figure}[htb]
        \includegraphics[width=0.92\columnwidth,clip]{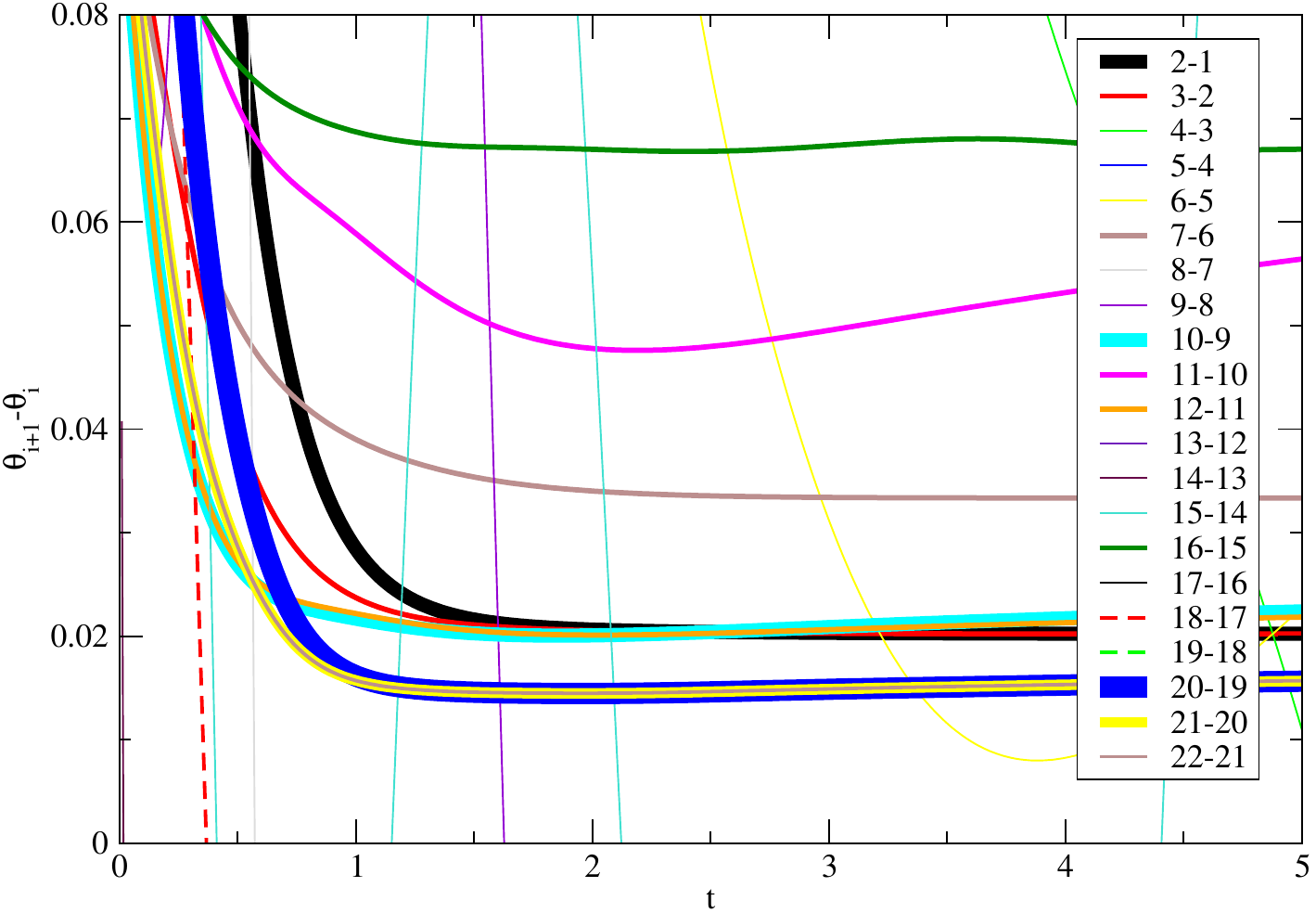}
        \caption{{Particular of Fig. 2.
          It shows in full detail all the ten phase differences $\theta_{i+1}-\theta_i$ associated
          to the ten links writable in the form $(i,i+1)$
          inside the groups of TE nodes of the graph of Fig. 1.
          It demonstrates the consistency of the general behavior 
          $\theta_{i+1}-\theta_i$ with the bounds established for example by Eq. (15), as well as by Eqs. (S11)-(S12) of SM:
          After an initial exponential drop, $\theta_{i+1}-\theta_i$ in each group reaches a smooth regime with a bounded value that is
          proportional to the spread of the group's frequencies and inversely proportional to the coupling $J$}.}
        \label{fig2_SM}
      \end{figure}  

{
  In conclusion, we analytically predict and confirm by simulations the local synchronization of oscillators with heterogeneous frequencies:
  groups of TE nodes with $k^{(\mathrm{OUT})}$ sufficiently smaller than $k^{(\mathrm{IN})}$ (depending on the group),
  synchronize and remain protected from the rest of the system, even under noise.
Plots of rotating numbers support the idea that TE groups act as independent pacemakers.
In fact, the number of convergence lines scales with the number of TE groups and, as the coupling $J$ increases,
these lines approach and eventually merge into one, indicating global synchronization, consistently with~\cite{ArenasPRL,Pecora}.
However, for any finite value of $J$, each group of TE oscillators satisfying the TC maintains its own synchronization status, regardless of the others.
This is precisely what is expected, \textit{e.g.}, in many biological systems, whose  
functionality relies on the coexistence of several independent pacemakers.
Our exact result could have broad applications since,
as shown in \cite{MacArthur}, in real-world networks,
the number of groups of FC TE nodes scales with the system size $N$, ranging from 20 \% to 98\%.}

{This work was supported by the National Council for Scientific and Technological Development (CNPq), Brazil, under the Universal Grant (process no. 409180/2023-8).}


\clearpage

\pagebreak
\widetext
\begin{center}
\textbf{\large Supplemental Material for \\
  Topologically protected synchronization in networks}
\end{center}
\setcounter{equation}{0}
\setcounter{figure}{0}
\setcounter{table}{0}
\setcounter{page}{1}
\makeatletter
\renewcommand{\theequation}{S\arabic{equation}}
\renewcommand{\thefigure}{S\arabic{figure}}
\renewcommand{\bibnumfmt}[1]{[S#1]}
\renewcommand{\citenumfont}[1]{S#1}

{
\section{Proof of Eq. (10) of MP}
The left inequality in Eq. (10) of MP is well known. As for the right inequality consider
the Laurent series for $\sin(\theta)$
\begin{eqnarray}
  \label{Simple}
&&  -\sin(\theta)=-\sum_{k=0}^\infty \frac{\theta^{2k+1}}{(2k+1)!}(-1)^k=-\theta+\frac{\theta^3}{3!}-\frac{\theta^5}{5!}+\frac{\theta^7}{7!}-\ldots.
\end{eqnarray}
Applying Eq. (\ref{Simple}) with $0\leq \theta<1$ we get
\begin{eqnarray}
  \label{Simple1}
&&  -\sin(\theta)\leq -\theta+\frac{\theta}{3!}+\frac{\theta}{7!}+\ldots=-\theta(1-\gamma).
\end{eqnarray}
}

\section{Proof that $\lambda<0$ and $0<\Theta_0<1$ imply $0<\Theta<1$, $\forall t>0$}
We limit the proof to the case of equal frequencies.
Let us start from Eq. (13) of MP that here we rewrite for convenience:
\begin{eqnarray}
\label{Rigor2SM}
&& {\Theta}_0 e^{-J \mu t} < {\Theta} < {\Theta}_0 e^{J \lambda t},  
\end{eqnarray}
which is valid for $t\in[0,t_1)$ under the hypothesis that $0<{\Theta}_0<1$.
Recall that $[0,t_1)$ is defined as the interval where the solution satisfies $0<\Theta(t)<1$.
  Let us observe that the Kuramoto-like model, Eq. (1) of MP, implies that, for any $i=1,\ldots,N$,
  both $\theta_i(t)$ and $\dot{\theta}_i(t)$ are continuous functions of time in the interval $[0,\infty)$
    (they are actually class $C^{\infty}[0\,+\infty)$
    but for us it suffices to observe that they are --- at least --- class $C^{1}[0\,+\infty)$).
      In turn, this implies the continuity of both the phase difference $\Theta(t)$ and its time derivative $\dot{\Theta}(t)$.

      Let us suppose that, by contradiction, $t_1<\infty$.
  This implies that it is either $\Theta(t_1)=1$ or $\Theta(t_1)=0$. We shall see in a moment that both possibilities have to be discarded.

  If we suppose the first possibility, $\Theta(t_1)=1$, from the upper bound in Eq. (\ref{Rigor2SM}) it follows that,
    for any $0<\epsilon<t_1$, by using $\lambda<0$, we have 
\begin{eqnarray}
  \label{Proof1}
  \Theta(t_1)-\Theta(t_1-\epsilon)> 1-\Theta_0 e^{J \lambda (t_1-\epsilon)}>1-\Theta_0>0.   
\end{eqnarray}
The above expression would imply that $\Theta(t)$ is discontinuous at $t_1$.   

If we suppose the second possibility, $\Theta(t_1)=0$, then, from the lower bound in Eq. (\ref{Rigor2SM}) it follows that,
for any $0<\epsilon<t_1$,
\begin{eqnarray}
  \label{Proof2}
\Theta(t_1)-\Theta(t_1-\epsilon)< -\Theta_0 e^{-J \mu (t_1-\epsilon)}.   
\end{eqnarray}
For $t_1$ finite, the above expression would imply that $\Theta(t)$ is discontinuous at $t_1$. The proof is complete.   
%

\section{Other examples of groups of topologically equivalent nodes}
Depending on the relation between $k^{(\mathrm{IN})}$ and $N'$, a group of $N'$ topologically equivalent (TE) nodes
can be realized in several ways. Fig. 1 of the main paper (MP) included cases with $N'=2, 3$ and $N'=4$.
Here, Fig. \ref{fig1_SM} shows all possible cases with for $N'=2$ to $5$.
The synchronization scenario and the topological condition (TC) run as follows:
\begin{itemize}
  \item When the nodes of the group have no connections among each other (Panels a, c, e, m), there is no stable attractor but, in the case of equal frequencies,
    the dynamics of such TE oscillators
    is governed by the same identical equation so that, over time, each oscillator tracks exactly the other
   \item On the base of the cases that we were able to calculate explicitly so far,
    when the nodes of the group form a fully connected (FC) graph (like Panels b, d, l, o), 
    the sufficient TC for synchronization of the group is given by $k^{(\mathrm{OUT})} \leq k^{(\mathrm{IN})}$ (Panels b, d. l) or
    $k^{(\mathrm{OUT})} < k^{(\mathrm{IN})}$ (Panel o), depending on the case.
   \item Similarly, when the group forms a disjoint union of $m$ FC graphs (Panels f, g, h),
    the sufficient TC for the individual synchronization of each component is
    $k^{(\mathrm{OUT})}\leq \min_{l\in\{1,\ldots,m\}} k_l^{(\mathrm{IN})}$, or $k^{(\mathrm{OUT})}\leq \min_{l\in\{1,\ldots,m\}} k_l^{(\mathrm{IN})}$, depending on the case, 
    where $k_l^{(\mathrm{IN})}$ is the internal degree
    of the $l$-th FC component.
    \item When the nodes of the group form instead regular polygons, unless special initial conditions are chosen,
    the chance for synchronization decreases or ceases. For example,
    for the case $N'=4$ (Panel i) $k^{(\mathrm{OUT})}$ can be at most 1, while for the case $N'=5$ (Panel n), $k^{(\mathrm{OUT})}$ must be 0. 
\end{itemize}    

\begin{figure}[h]
  \centering
  \includegraphics[width=0.10\columnwidth,clip]{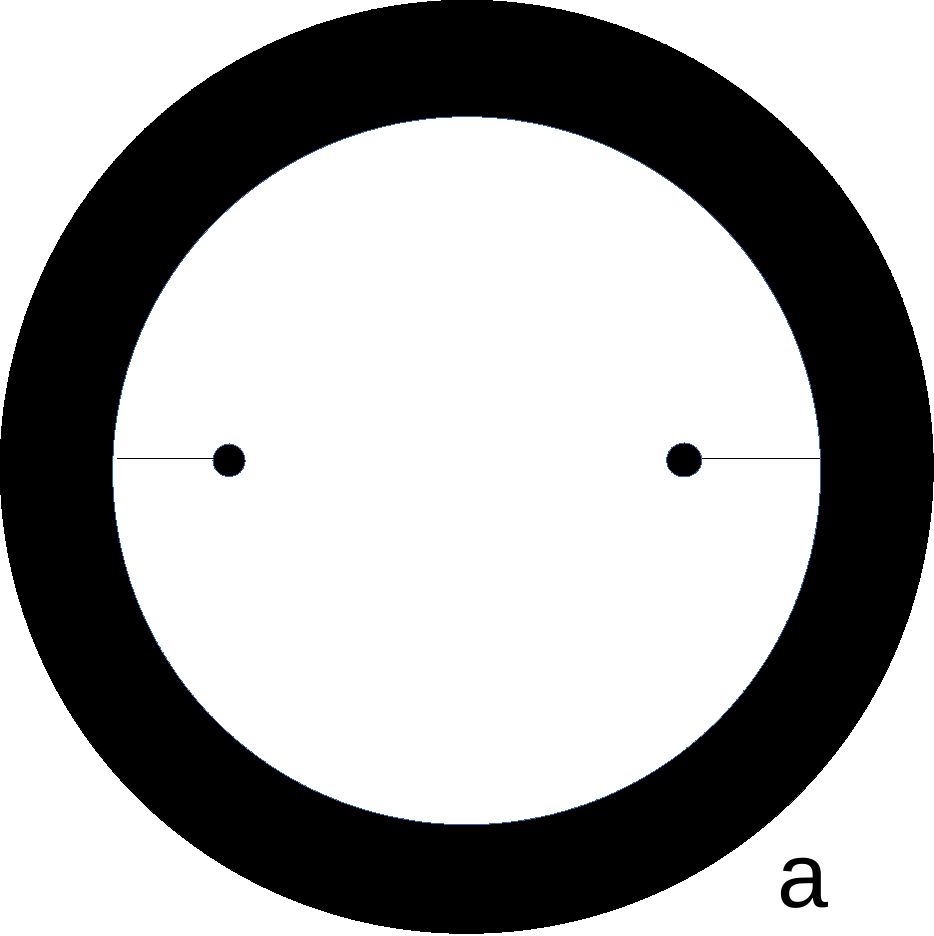}
  \includegraphics[width=0.10\columnwidth,clip]{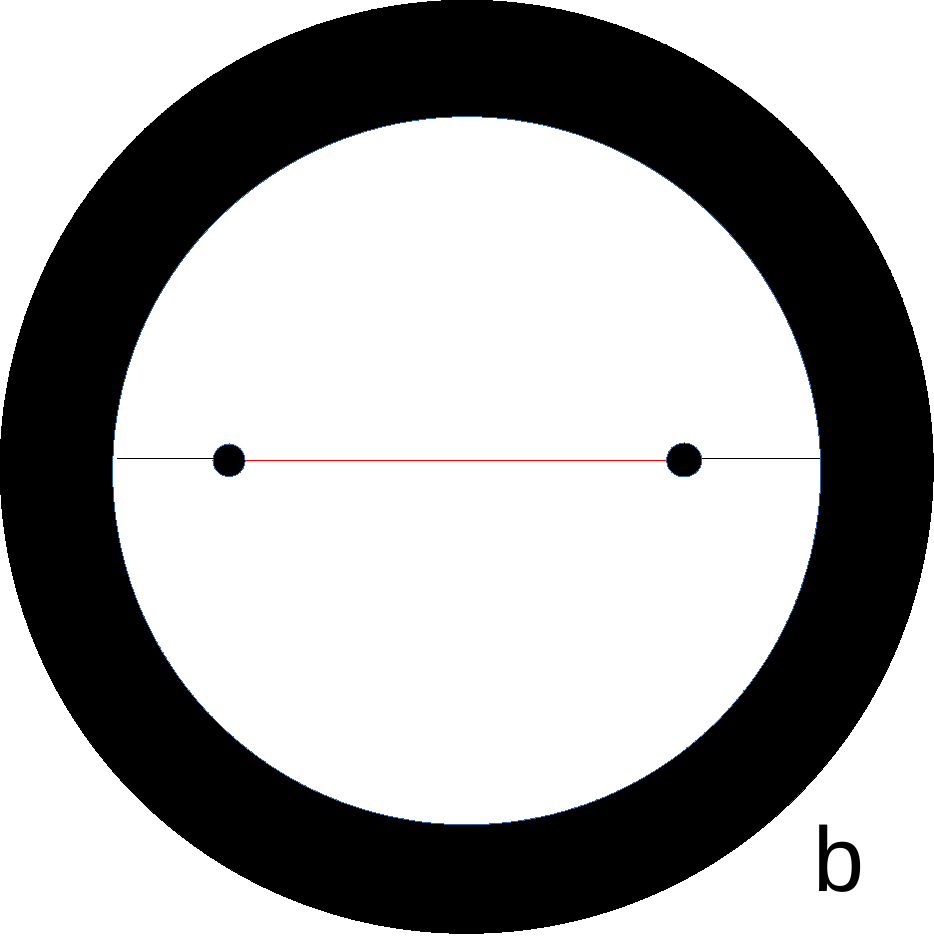}\\
  \includegraphics[width=0.10\columnwidth,clip]{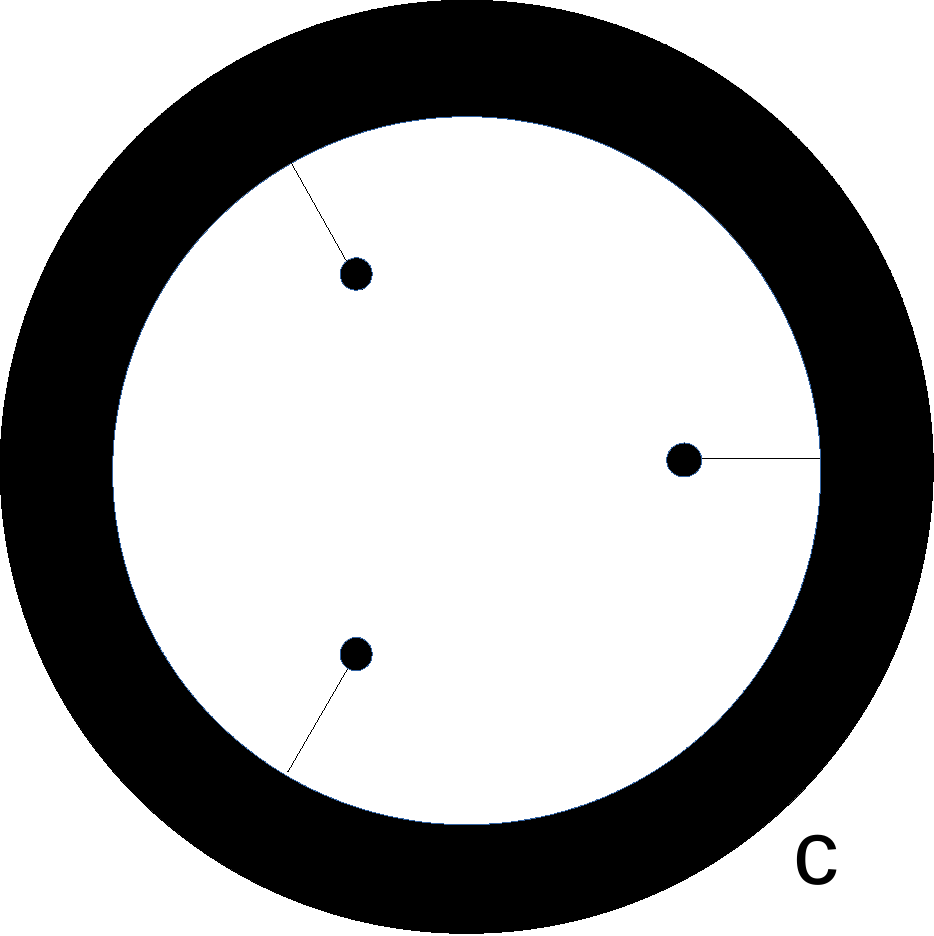}
  \includegraphics[width=0.10\columnwidth,clip]{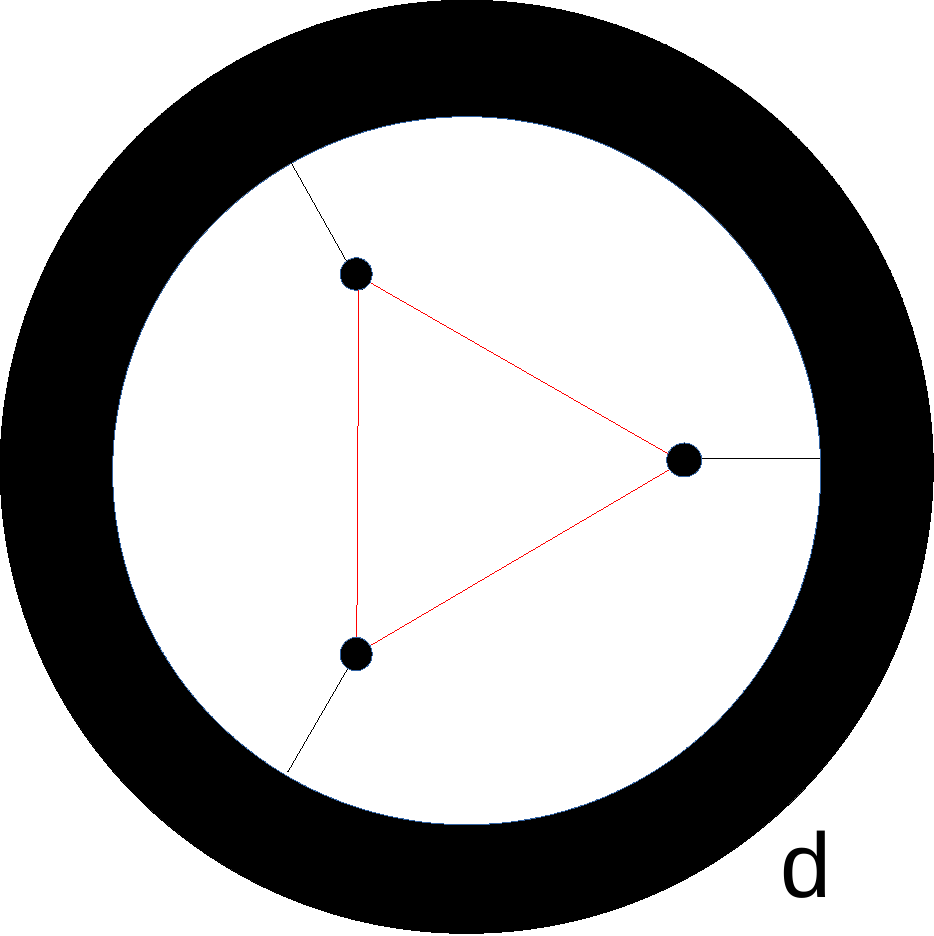}\\
    \includegraphics[width=0.10\columnwidth,clip]{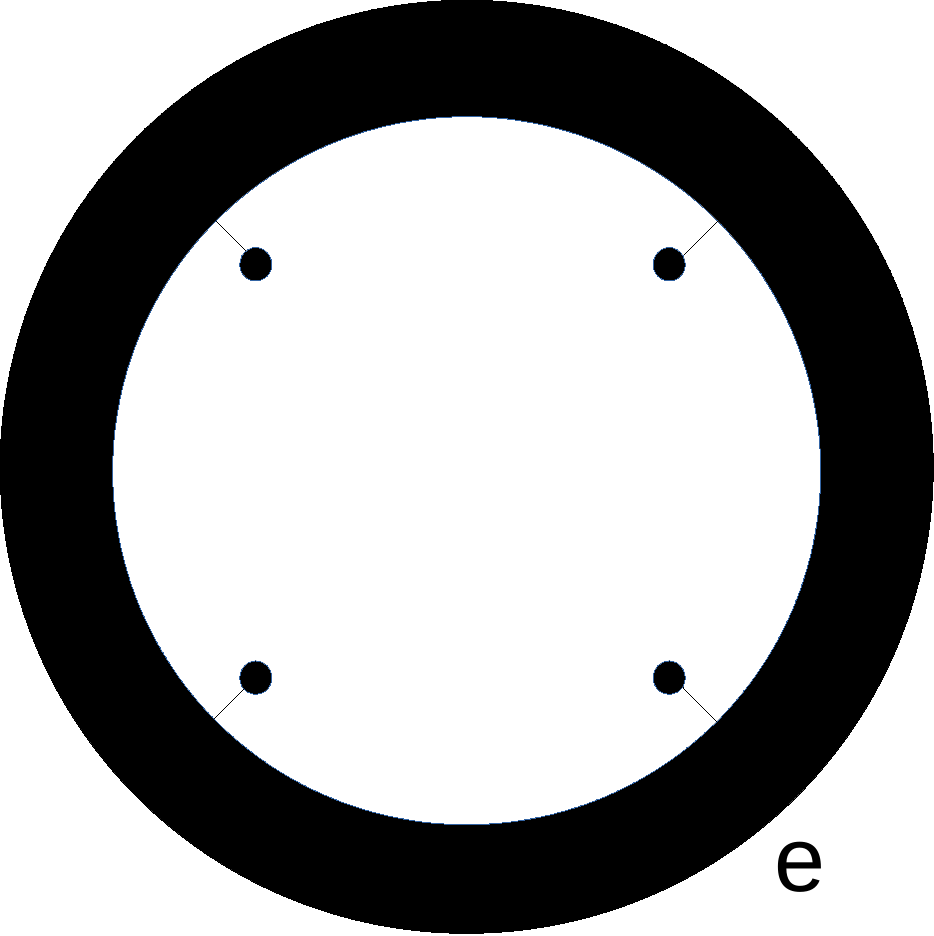}
    \includegraphics[width=0.10\columnwidth,clip]{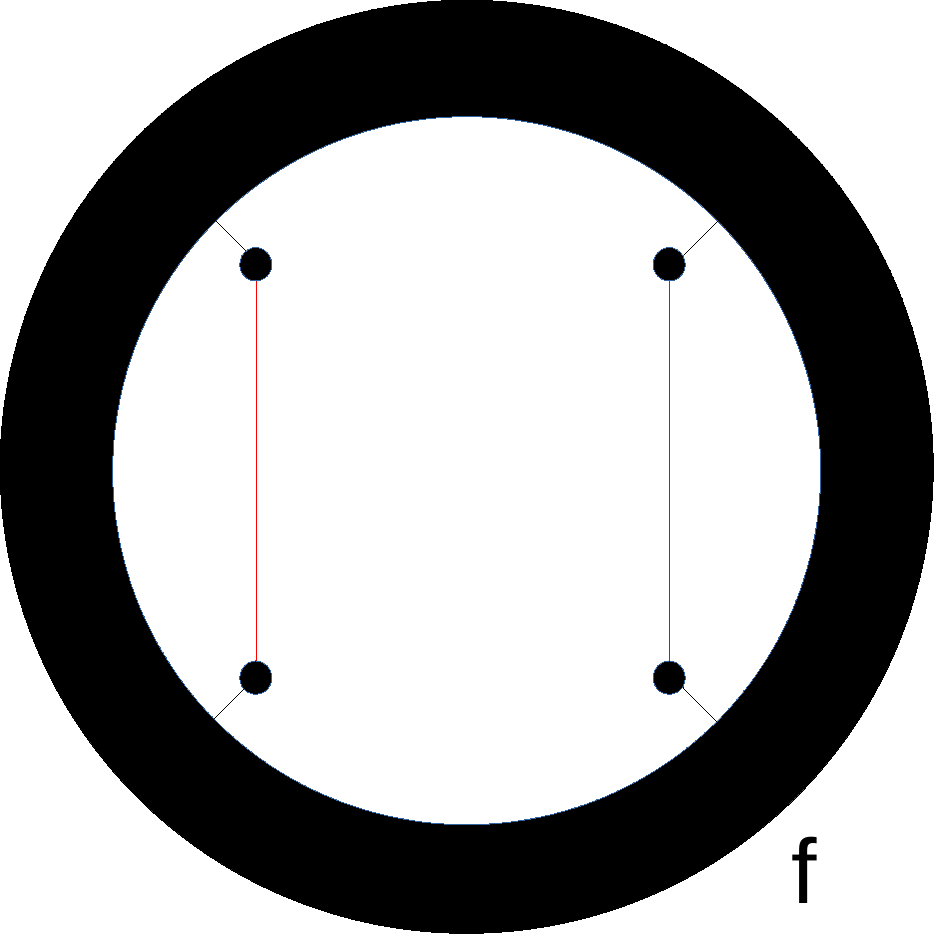}
    \includegraphics[width=0.10\columnwidth,clip]{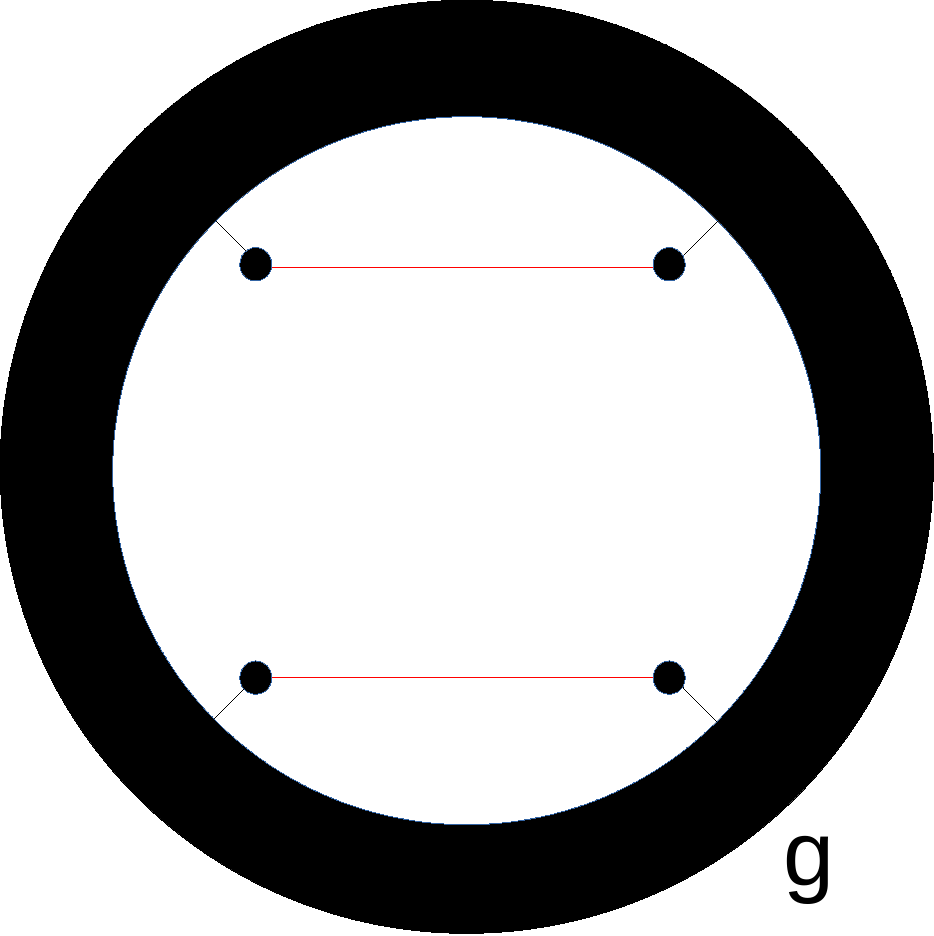}
    \includegraphics[width=0.10\columnwidth,clip]{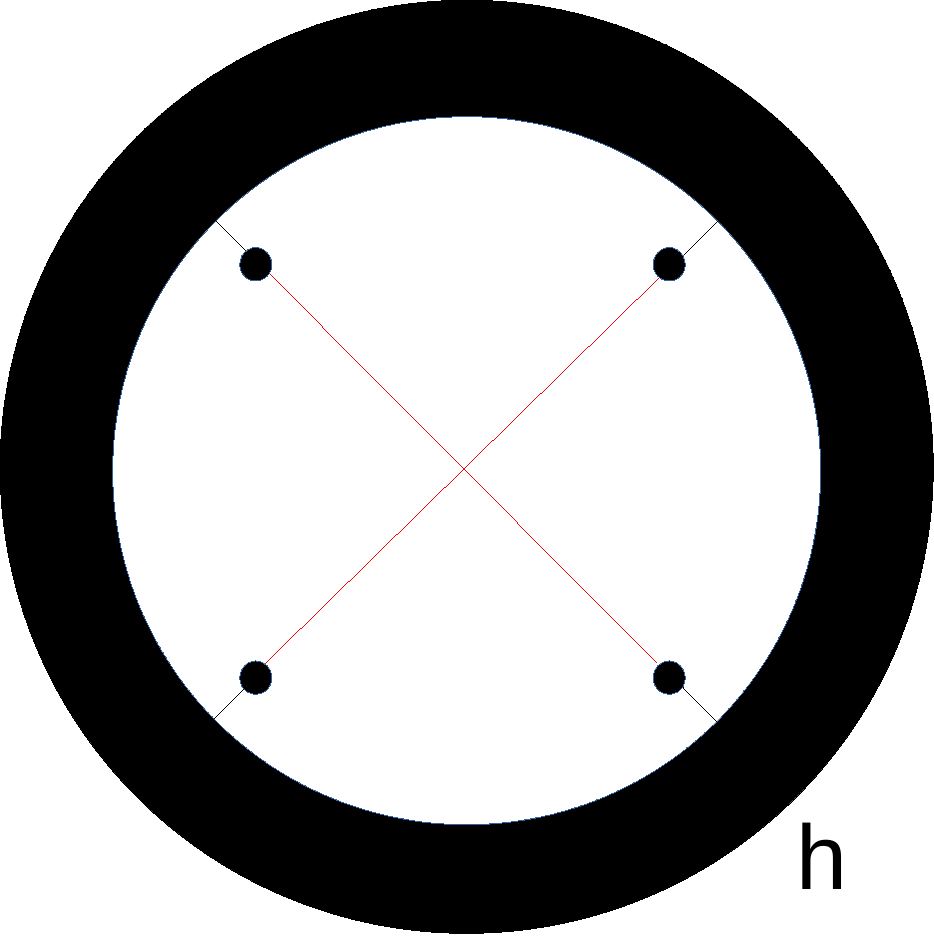}
    \includegraphics[width=0.10\columnwidth,clip]{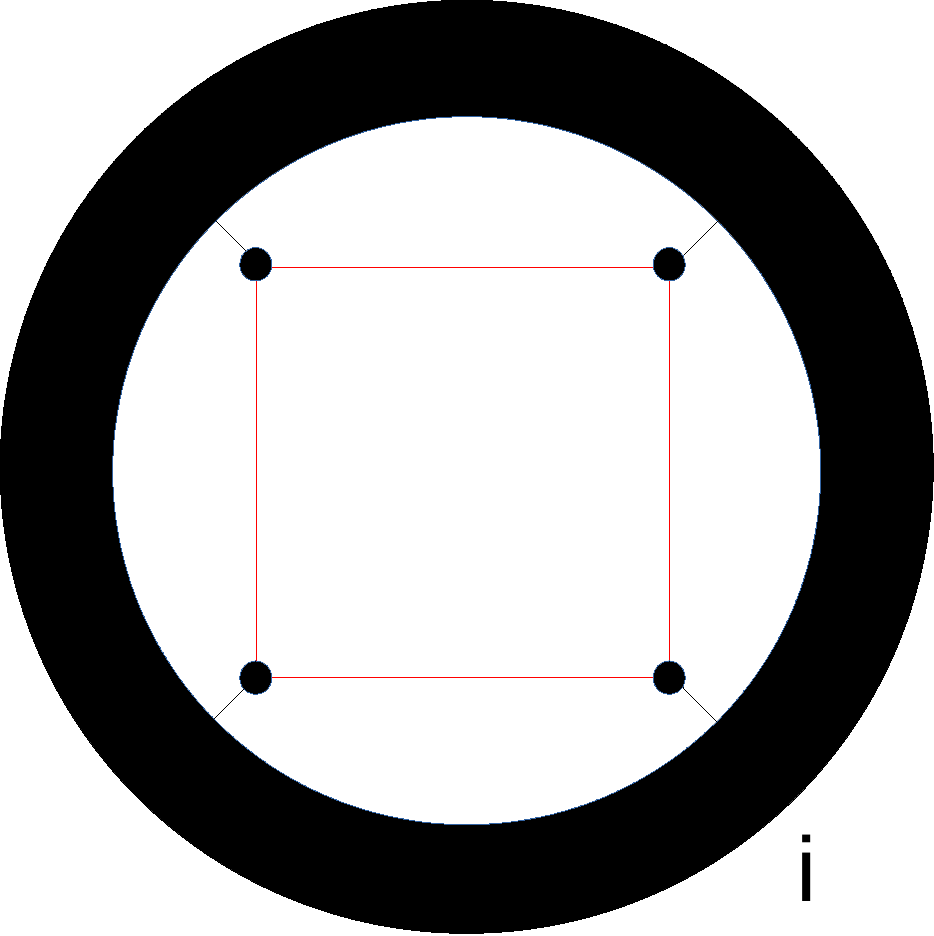}
    \includegraphics[width=0.10\columnwidth,clip]{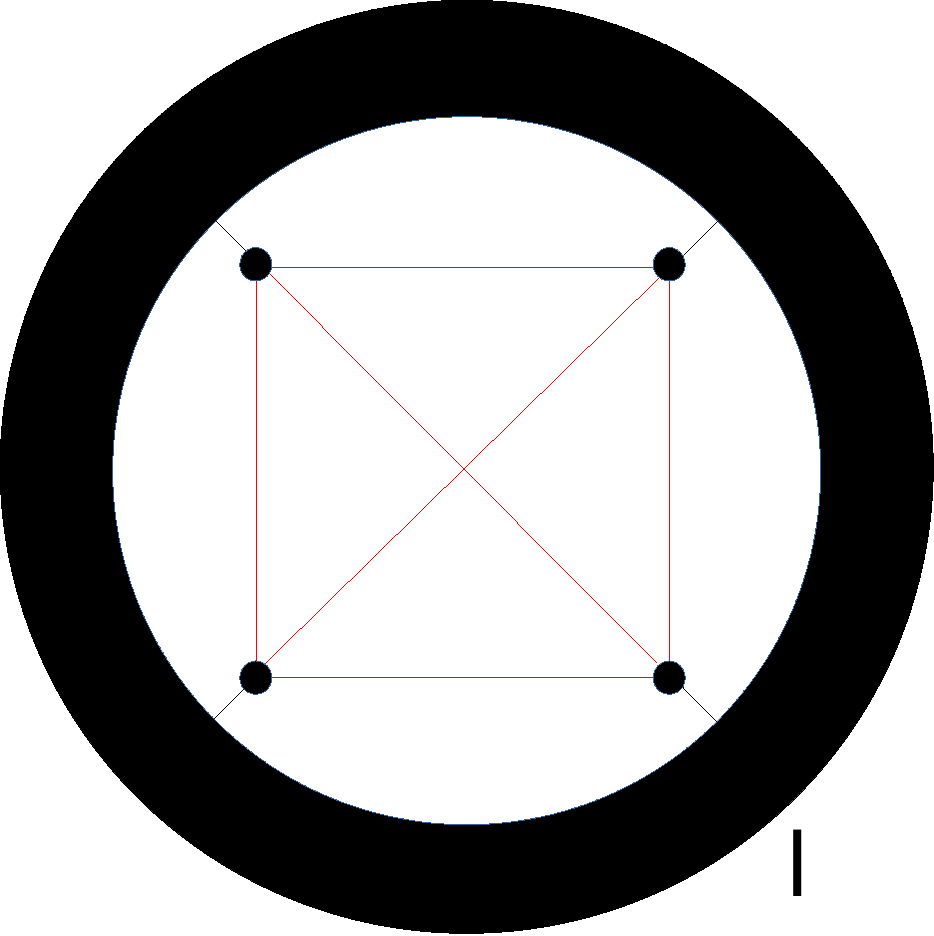}\\
    \includegraphics[width=0.10\columnwidth,clip]{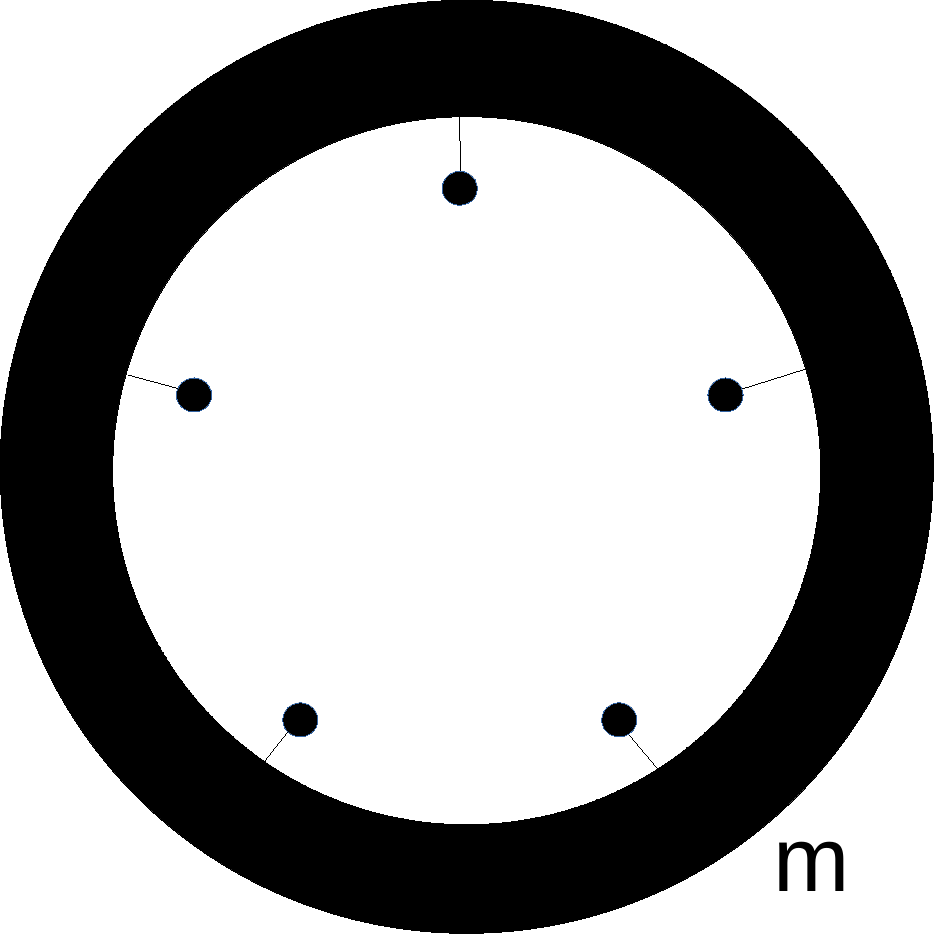}
    \includegraphics[width=0.10\columnwidth,clip]{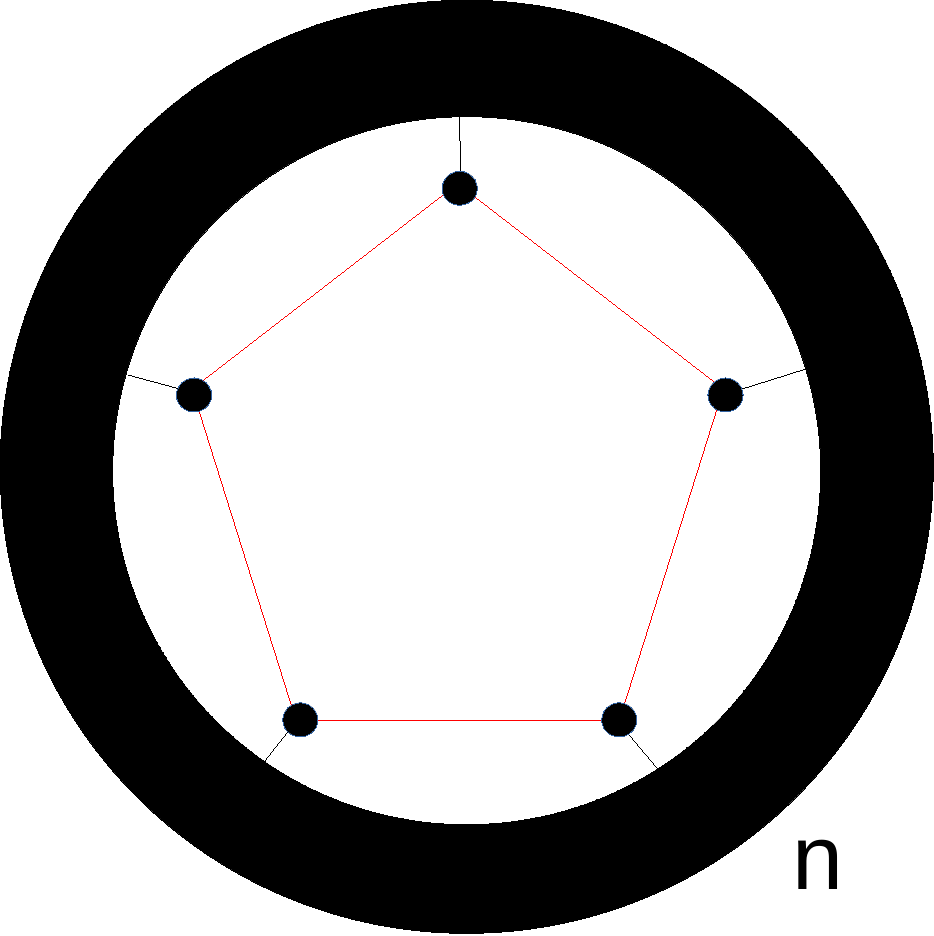}
    \includegraphics[width=0.10\columnwidth,clip]{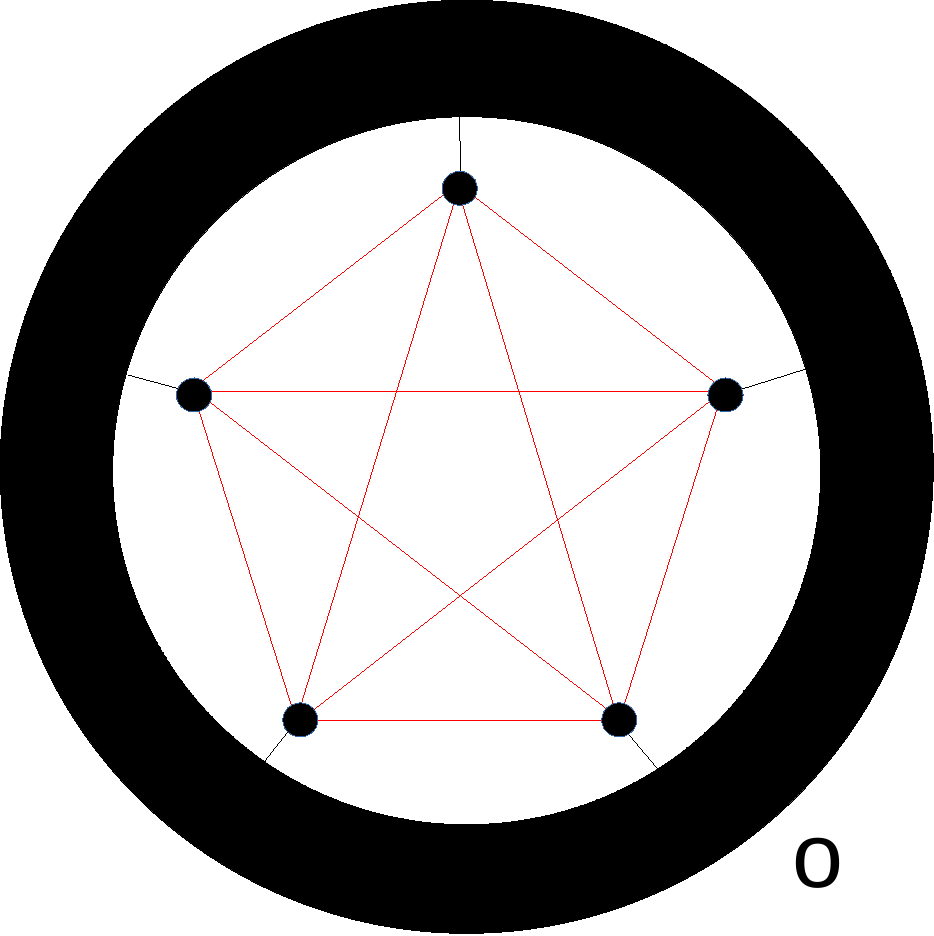}
  \caption{
    All possible groups of $N'$ TE nodes for $N'=2$ to $5$ (top to bottom). Here we focus only on the concept of topological equivalence within the group,
    while the topological
    equivalence with respect to the neighbors of the group --- indicated by the surrounding black corona --- is assumed.
    }
  \label{fig1_SM}
\end{figure}  

\section{Derivation of the TC for $N'>2$}
In order to analyze the synchronization, it useful to introduce some notation and point out a few crucial points.
Given $N'$ TE nodes with indices $1,2,\ldots,N'$, we indicate their corresponding phase differences by ${\Theta}_{i,j}=\theta_i-\theta_j$ and frequencies differences by ${\Omega}_{i,j}=\omega_i-\omega_j$.
Moreover, we generalize the definition introduced earlier for the difference between two ``external fields'':
$h_{i,j}'(\theta^{(*)}_{i,j})$ is such that $h_{i,j}'(\theta^{(*)}_{i,j}){\Theta}_{i,j}=h(\theta_i)-h(\theta_j)$.
To keep the notation lighter, however, for what follows, the dependence of $h_{i,j}'(\theta^{(*)}_{i,j})$ on $\theta^{(*)}_{i,j}$ is unimportant and will be left understood.
Observe that ${\Theta}_{j,i}=-{\Theta}_{i,j}$ and that the variables $\{{\Theta}_{i,j}\}$ with $j>i$, are not all independent. For example, for $N'=3$ we have the constrain
${\Theta}_{3,1}={\Theta}_{3,2}+{\Theta}_{{2,1}}$.
In general, given $N'$, we can always write a system of $N'-1$ independent Ado's involving $N'-1$ independent variables.
We can always assume that the initial conditions ${\Theta}_{i,j}(0)$ with $j>i$ are all positive. If it occurs that any of them is not positive, as \textit{e.g.} ${\Theta}_{2,1}(0)$, it will be enough to
study the ODE involving ${\Theta}_{1,2}$ instead of ${\Theta}_{2,1}$.
We shall also assume $0<{\Theta}_{i,j}(0)<1/(N'-1)$ so that, at small enough times, we also have $0<{\Theta}_{i,j}(t)<1/(N'-1)$. Our general strategy is to
use the bounds $|h_{i,j}'|\leq k^{(\mathrm{OUT})}$; $-\sin(\Theta)<\Theta(1-\gamma)$, with $\gamma$ given by Eq. (11) of MP; 
and the bound $\sin({\Theta})\leq{\Theta}$.
If the found bounding solution keeps satisfying the above assumptions for any time $t$, the procedure is consistent.

The resulting bounding ODE can be written vectorially:
\begin{align}
  \label{bbb}
\dot{\bm{{\Theta}}}\leq \bm{{\Omega}} + J \bm{B}\cdot \bm{{\Theta}},
\end{align}
where $\bm{{\Theta}}=({\Theta}_{2,1},{\Theta}_{3,2},\ldots,{\Theta}_{N',N'-1})^T$, $\bm{{\Omega}}=({\Omega}_{2,1},{\Omega}_{3,2},\ldots,{\Omega}_{N',N'-1})^T$,
and $\bm{B}$ is a $(N'-1)\times (N'-1)$ suitable matrix that depends on the parameters
$N'$, $k^{(\mathrm{OUT})}$, $k^{(\mathrm{IN})}$, $\gamma$, and on the specific way in which the TE nodes are arranged among each other.
Let $\lambda_i$ and $\bm{u}_i$ be the eigenvalues and normalized eigenvectors of $\bm{B}$, respectively. The coefficients $c_i(t)=\bm{{\Theta}}\cdot \bm{u}_i$
satisfy the decoupled system $d c_i(t)/dt \leq \bm{{\Omega}}\cdot \bm{u}_i+J \lambda_i c_i(t)$. As a consequence, each component of the vector $\bm{{\Theta}}$ remains bounded
if all the eigenvalues of $\bm{B}$ are negative.

Below we consider FC cases and and then regular polygons. 

\subsection{$N'=3$ FC (Panel d of Fig. \ref{fig1_SM})}
Here we show the bounding solution in full detail.
We have the following ODE system
\begin{align}
  \frac{\dot{{\Theta}}_{2,1}}{J}\leq
  \frac{{\Omega}_{2,1}}{J}-2\sin\left({\Theta}_{2,1}\right)+\sin\left({\Theta}_{3,2}\right)-\sin\left({\Theta}_{3,1}\right)+h_{2,1}'{\Theta}_{2,1},\nonumber \\
  \frac{\dot{{\Theta}}_{3,2}}{J}\leq
  \frac{{\Omega}_{3,2}}{J}-2\sin\left({\Theta}_{3,2}\right)+\sin\left({\Theta}_{2,1}\right)-\sin\left({\Theta}_{3,1}\right)+h_{3,2}'{\Theta}_{3,2}.\nonumber,
\end{align}
to be solved with the constrain
\begin{align}
{\Theta}_{3,1}={\Theta}_{3,2}+{\Theta}_{2,1}.
\end{align}
By using the three bounds explained previously we get the following matrix
\begin{align}
  \bm{B}=
  \begin{bmatrix}
    k^{(\mathrm{OUT})}-3(1-\gamma)       & \gamma \\
    \gamma              & k^{(\mathrm{OUT})}-3(1-\gamma)
  \end{bmatrix},
\end{align}
to be used in Eq. (\ref{bbb}) with the vectors
$\bm{{\Theta}}=({\Theta}_{2,1},{\Theta}_{3,2})^T$ and $\bm{{\Omega}}=({\Omega}_{2,1},{\Omega}_{3,2})^T$.
The eigenvalues of $\bm{B}$ are
\begin{align}
  \lambda_1=k^{(\mathrm{OUT})}-3 +4\gamma,\\
  \lambda_2=k^{(\mathrm{OUT})}-3 +2\gamma,
\end{align}
with normalized eigenvectors $\bm{u}_1=(1,1)^T/\sqrt{2}$ and $\bm{u}_2=(1,-1)^T/\sqrt{2}$, respectively.

We conclude that, for a group of $N'=3$ TE FC nodes, a sufficient TC for synchronization is $k^{(\mathrm{OUT})}-3 +4\gamma<0$, which, on observing that
$0<4\gamma<1$, amounts
to $k^{(\mathrm{OUT})}\leq 2$, or $k^{(\mathrm{OUT})}\leq k^{(\mathrm{IN})}$, where we have made use of the fact that, here, $k^{(\mathrm{IN})}=2$.
By integrating the two normal modes $c_1(t)$ and $c_2(t)$, we obtain the following bounding solution
  \begin{align}
    \label{aaa3}
          {\Theta}_{2,1}\leq \frac{\Omega_{2,1}+\Omega_{3,2}}{2J|\lambda_1|}+\frac{\Omega_{2,1}-\Omega_{3,2}}{2J|\lambda_2|}
          +\left(-\frac{\Omega_{2,1}+\Omega_{3,2}}{2J|\lambda_1|}+\frac{c_1(0)}{\sqrt{2}}\right)\exp\left(-J|\lambda_1|t\right)+
          \left(-\frac{\Omega_{2,1}-\Omega_{3,2}}{2J|\lambda_2|}+\frac{c_2(0)}{\sqrt{2}}\right)\exp\left(-J|\lambda_2|t\right) \\
          \label{bbb3}
          {\Theta}_{3,2}\leq \frac{\Omega_{2,1}+\Omega_{3,2}}{2J|\lambda_1|}-\frac{\Omega_{2,1}-\Omega_{3,2}}{2J|\lambda_2|}
          +\left(-\frac{\Omega_{2,1}+\Omega_{3,2}}{2J|\lambda_1|}+\frac{c_1(0)}{\sqrt{2}}\right)\exp\left(-J|\lambda_1|t\right)-
          \left(-\frac{\Omega_{2,1}-\Omega_{3,2}}{2J|\lambda_2|}+\frac{c_2(0)}{\sqrt{2}}\right)\exp\left(-J|\lambda_2|t\right),
\end{align}
where $c_1(0)=\bm{{\Theta}}(t=0)\cdot \bm{u}_1$ and $c_2(0)=\bm{{\Theta}}(t=0)\cdot \bm{u}_2$ are the projections of the initial conditions onto the eigenvectors.
Note that, for $J$ sufficiently large, all the above procedure turns out to be consistent with the required bound $0<{\Theta}_{i,j}<1/2$ for any $t$.
More precisely, by using $|\lambda_i|>1/2$, from Eqs. (\ref{aaa3})-(\ref{bbb3}) it follows that a sufficient condition for ${\Theta}_{i,j}<1/2$ is $J>4 \max_{i,j}|{\Omega}_{i,j}|$.
On the other hand, by using $|\lambda_2|>|\lambda_1|$ together with $\Omega_{2,1}\geq 0$ and $\Omega_{3,2}\geq 0$, it follows also that $0<{\Theta}_{i,j}$.

\subsection{$N'=4$ FC (Panel l of Fig. \ref{fig1_SM})}
Here, the ODE system must be solved with the following constrains
\begin{align}
  &{\Theta}_{4,1}={\Theta}_{4,2}+{\Theta}_{3,2}+{\Theta}_{2,1},\\
  &{\Theta}_{4,2}={\Theta}_{4,3}+{\Theta}_{3,2},\\
  &{\Theta}_{3,1}={\Theta}_{3,2}+{\Theta}_{2,1}.
\end{align}
By using the three bounds we obtain the following matrix
\begin{align}
  \bm{B}=
  \begin{bmatrix}
    k^{(\mathrm{OUT})}-4(1-\gamma)       & 2\gamma & \gamma \\
    \gamma              & k^{(\mathrm{OUT})}-5(1-\gamma) & \gamma\\
    \gamma              & 2\gamma   & k^{(\mathrm{OUT})}-4(1-\gamma)
  \end{bmatrix}.
\end{align}
The eigenvalues of $\bm{B}$ are
\begin{align}
  &\lambda_1=k^{(\mathrm{OUT})}-4 +3\gamma,\\
  &\lambda_2=k^{(\mathrm{OUT})}-\frac{9}{2} +5\gamma-\frac{1}{2}\sqrt{1+16\gamma^2},\\
  &\lambda_3=k^{(\mathrm{OUT})}-\frac{9}{2} +5\gamma+\frac{1}{2}\sqrt{1+16\gamma^2}.
\end{align}
The largest of these three eigenvalues is $\lambda_3$, which, by using the fact that here is $k^{(\mathrm{IN})}=3$, and the explicit value of $\gamma=0.1668651044\ldots$, gives
\begin{align}
  \lambda_3=k^{(\mathrm{OUT})}-k^{(\mathrm{IN})}-\delta, \quad \delta\simeq 0.0646.
\end{align}
We conclude that for a group of $N'=4$ FC nodes, a sufficient TC for synchronization is $k^{(\mathrm{OUT})}\leq k^{(\mathrm{IN})}$.

\subsection{$N'=5$ FC (Panel o of Fig. \ref{fig1_SM})}
Here, the ODE system must be solved with the following constrains
\begin{align}
  &{\Theta}_{5,1}={\Theta}_{5,4}+{\Theta}_{4,3}+{\Theta}_{3,2}+{\Theta}_{2,1},\\
  &{\Theta}_{5,2}={\Theta}_{5,4}+{\Theta}_{4,3}+{\Theta}_{3,2},\\
  &{\Theta}_{4,1}={\Theta}_{4,3}+{\Theta}_{3,2}+{\Theta}_{2,1},\\
  &{\Theta}_{4,2}={\Theta}_{4,3}+{\Theta}_{3,2},\\
  &{\Theta}_{3,1}={\Theta}_{3,2}+{\Theta}_{2,1}.
\end{align}
By using the three bounds we obtain the following matrix
\begin{align}
  \bm{B}=
  \begin{bmatrix}
    k^{(\mathrm{OUT})}-5(1-\gamma)       & 3\gamma & 2\gamma & \gamma \\
    \gamma        & k^{(\mathrm{OUT})}-5(1-\gamma) & 2\gamma & \gamma\\
    \gamma        & 2\gamma & k^{(\mathrm{OUT})}-5(1-\gamma) &\gamma\\
    \gamma        & 2\gamma & 3\gamma & k^{(\mathrm{OUT})}-5(1-\gamma)
  \end{bmatrix}.
\end{align}
The eigenvalues of $\bm{B}$ are
\begin{align}
  &\lambda_1=k^{(\mathrm{OUT})}-5 +3\gamma,\\
  &\lambda_2=k^{(\mathrm{OUT})}-5 +4\gamma,\\
  &\lambda_3=k^{(\mathrm{OUT})}-5(1-\gamma)+\frac{3}{2}\gamma -\frac{1}{2}\sqrt{41}\gamma,\\
  &\lambda_4=k^{(\mathrm{OUT})}-5(1-\gamma)+\frac{3}{2}\gamma +\frac{1}{2}\sqrt{41}\gamma.
\end{align}
The largest of these eigenvalues is $\lambda_4$, which, by using the fact that here is $k^{(\mathrm{IN})}=4$, and the explicit value of $\gamma$, gives
\begin{align}
  \lambda_4=k^{(\mathrm{OUT})}-k^{(\mathrm{IN})}+\delta', \quad \delta'\simeq 0.51214.
\end{align}
We conclude that for a group of $N'=5$ FC nodes, a sufficient TC for synchronization is $k^{(\mathrm{OUT})}\leq k^{(\mathrm{IN})}-1$.

\subsection{General FC case}
Finding a bound for the largest eigenvalue of $\bm{B}$ in the general FC case remains a formidable task.
In general, we expect that the maximal eigenvalue $\bm{B}$ scales as
$k^{(\mathrm{OUT})}-N'(1-\gamma)+\alpha N'=k^{(\mathrm{OUT})}-\left(k^{(\mathrm{IN})}+1\right)\left(\gamma+\alpha\right)$,
where $\alpha=\mathop{O}(1)$. This guess needs further demanding analysis.

\subsection{Regular polygon $N'=4$ (Panel i of Fig. 1)}
As shown in Fig. \ref{fig1_SM}, $N'=4$ provides the first situation where the $N'$ TE nodes can be arranged in more than one way. 
Besides the FC case, where $k^{(\mathrm{IN})}=3$, there are the configurations where $k^{(\mathrm{IN})}=1$ or $k^{(\mathrm{IN})}=2$.
In the former case, the subgraph of the $N'$ TE nodes with their intra-links is a disconnected one and corresponds to the union of two disconnected pairs of TE nodes,
for which we have already seen the TC (for example ${\Theta}_{1,2}=0$ and ${\Theta}_{3,4}=0$): $k^{(\mathrm{OUT})}\leq k^{(\mathrm{IN})}$.
In the latter case, the four TE nodes form a square and 
the obtained ODE system is no longer totally decoupled.
After using the constrain $\Theta_{4,1}=\Theta_{4,3}+\Theta_{3,2}+\Theta_{2,1}$,
for the bounding ODE system of the independent variables ${\Theta}_{2,1},{\Theta}_{3,2},{\Theta}_{4,3}$,
we get following matrix
\begin{eqnarray}
  \bm{B}=
  \begin{bmatrix}
    k^{(\mathrm{OUT})}-3(1-\gamma)       & \gamma & \gamma-1 \\
    1      & k^{(\mathrm{OUT})}-2(1-\gamma) &  1 \\
    \gamma-1              & \gamma & k^{(\mathrm{OUT})}-3(1-\gamma)
\end{bmatrix}.
\end{eqnarray}
The eigenvalues of $\bm{B}$ are
\begin{align}
  &\lambda_1=k^{(\mathrm{OUT})}-2 +2\gamma,\\
  &\lambda_1=k^{(\mathrm{OUT})}-3 +3\gamma-\sqrt{1+\gamma^2},\\
  &\lambda_1=k^{(\mathrm{OUT})}-3 +3\gamma+\sqrt{1+\gamma^2}.
\end{align}
The largest of these eigenvalues is $\lambda_3$, which, by using the fact that here is $k^{(\mathrm{IN})}=2$, and the explicit value of $\gamma$, gives
\begin{align}
  \lambda_3=k^{(\mathrm{OUT})}-2+\delta=k^{(\mathrm{OUT})}-k^{(\mathrm{IN})}+\delta'', \quad \delta''\simeq 0.5144.
\end{align}
We conclude that for this group, a sufficient TC for synchronization is $k^{(\mathrm{OUT})}\leq 1$.

\subsection{Regular polygon $N'=5$ (Panel n of Fig. 1)}
After using the constrain $\Theta_{5,1}=\Theta_{5,4}+\Theta_{4,3}+\Theta_{3,2}+\Theta_{2,1}$, 
for the bounding ODE system of the independent variables ${\Theta}_{2,1},{\Theta}_{3,2},{\Theta}_{4,3},{\Theta}_{5,4}$,
we get following matrix
\begin{eqnarray}
  \bm{B}=
  \begin{bmatrix}
    k^{(\mathrm{OUT})}-3(1-\gamma)  & \gamma & \gamma-1 & \gamma-1 \\
    1            & k^{(\mathrm{OUT})}-2(1-\gamma) &  1  &  0 \\
    0 & 1        & k^{(\mathrm{OUT})}-2(1-\gamma) & 1\\
    \gamma-1 & \gamma-1      & \gamma & k^{(\mathrm{OUT})}-3(1-\gamma)
\end{bmatrix}.
\end{eqnarray}
The eigenvalues of $\bm{B}$ are
\begin{align}
  &\lambda_1=k^{(\mathrm{OUT})}-\frac{5}{2}-\frac{\sqrt{5}}{2} +2\gamma,\\
  &\lambda_2=k^{(\mathrm{OUT})}-\frac{5}{2}+\frac{\sqrt{5}}{2} +2\gamma,\\
  &\lambda_3=k^{(\mathrm{OUT})}-\frac{5}{2}+3\gamma-\frac{\sqrt{5-4\gamma+4\gamma^2}}{2},\\
  &\lambda_4=k^{(\mathrm{OUT})}-\frac{5}{2}+3\gamma+\frac{\sqrt{5-4\gamma+4\gamma^2}}{2}.
\end{align}
The largest of these eigenvalues is $\lambda_4$, which, by using the explicit value of $\gamma$, gives
\begin{align}
  \lambda_4=k^{(\mathrm{OUT})}-\delta''', \quad \delta'''\simeq 0.9453.
\end{align}
We conclude that for this group, a sufficient TC for synchronization is $k^{(\mathrm{OUT})}=0$.

\section{Simulations}
The simulations reported in MP have been realized by means of the Runge-Kutta method\cite{Numerical_Recipes_SM} at fourth-order with step parameter $h=0.01$.
Concerning the precision of the method,
each choice of initial conditions has been accepted or not with the following criterion. If for natural frequencies all equal and $J=1$ the simulation
provides global synchronization (which is an exact result) the choice is accepted, otherwise is rejected.

As for the simulations reported in Fig. 2 of MP, the initial conditions and the natural frequencies, respectively, are:\\
      $\theta_{1}=1.2;
      \theta_{2}=1.5;
      \theta_{3}=1.6;
      \theta_{4}=3.2;
      \theta_{5}=3.8;
      \theta_{6}=4.1;
      \theta_{7}=4.2;
      \theta_{8}=5.3;
      \theta_{9}=5.4;
      \theta_{10}=5.5;
      \theta_{11}=5.6;
      \theta_{12}=5.7;
      \theta_{13}=6.5;
      \theta_{14}=6.6;
      \theta_{15}=7.4;
      \theta_{16}=7.5;
      \theta_{17}=7.6;
      \theta_{18}=7.9;
      \theta_{19}=8.0;
      \theta_{20}=8.2;
      \theta_{21}=8.3;
      \theta_{22}=8.4$,
and
    $\omega_1=2.1,\omega_2=2.2,\omega_3=2.3,\omega_4=4.5,
   \omega_5=3.5,\omega_6=2.6,\omega_7=2.7,\omega_8=1.1,
       \omega_9=1.2,\omega_{10}=1.3,\omega_{11}=1.4,\omega_{12}=1.5,
       \omega_{13}=2.5,\omega_{14}=0.8,\omega_{15}=0.5,\omega_{16}=0.7,
       \omega_{17}=1.5,\omega_{18}=0.8,
      \omega_{19}=3.0,\omega_{20}=3.1,\omega_{21}=3.2,\omega_{22}=3.3$.

      Here, Fig. \ref{fig2_SM} {shows the same Fig. 2 of MP in further detail by means of four increasing
        zooms.}
      In Fig. \ref{fig3_SM} we show also the rotating numbers for the same system of Fig. 2 of MP.

      \begin{figure}[htb]
        \centering
        \includegraphics[width=0.49\columnwidth,clip]{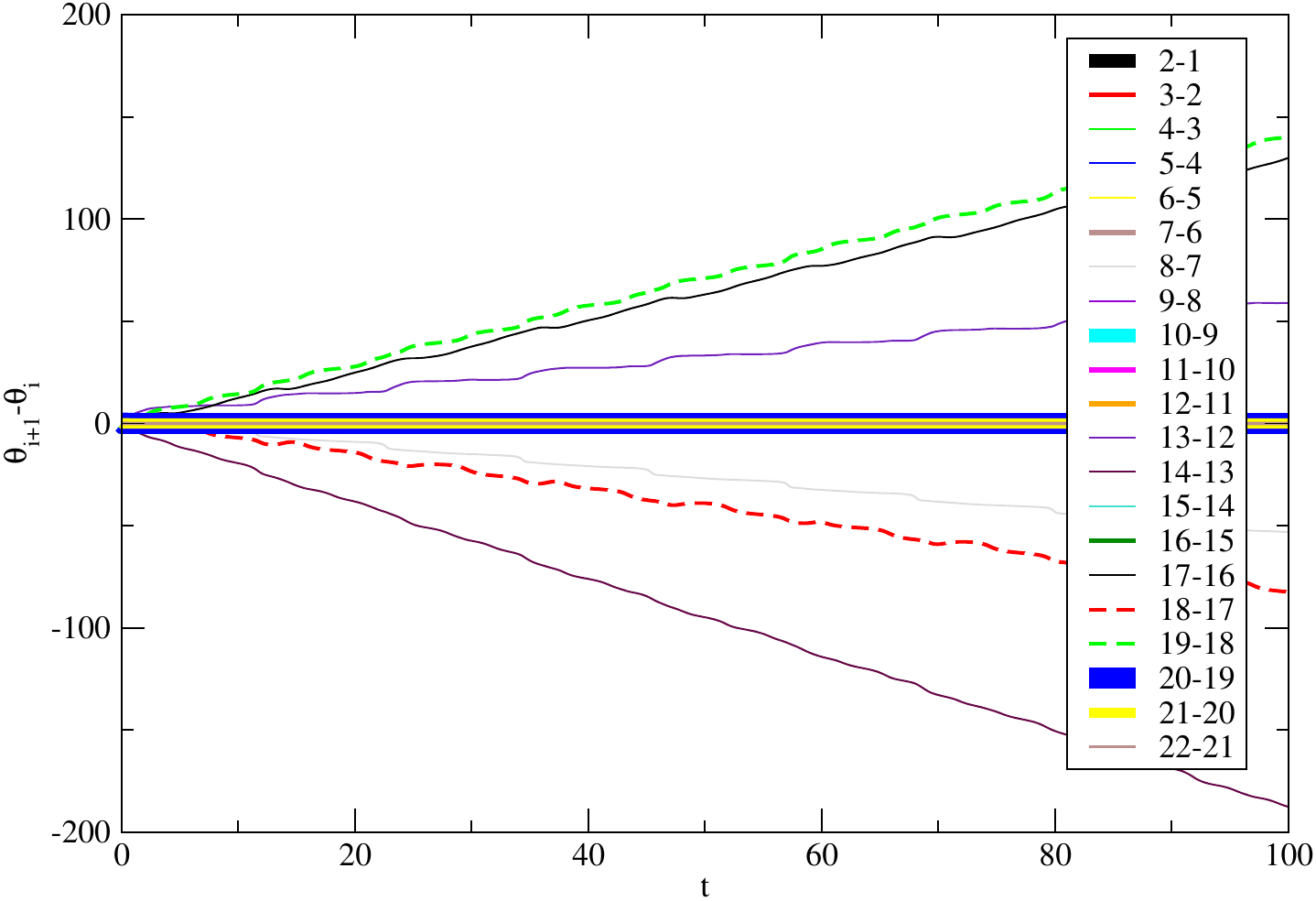}
        \includegraphics[width=0.49\columnwidth,clip]{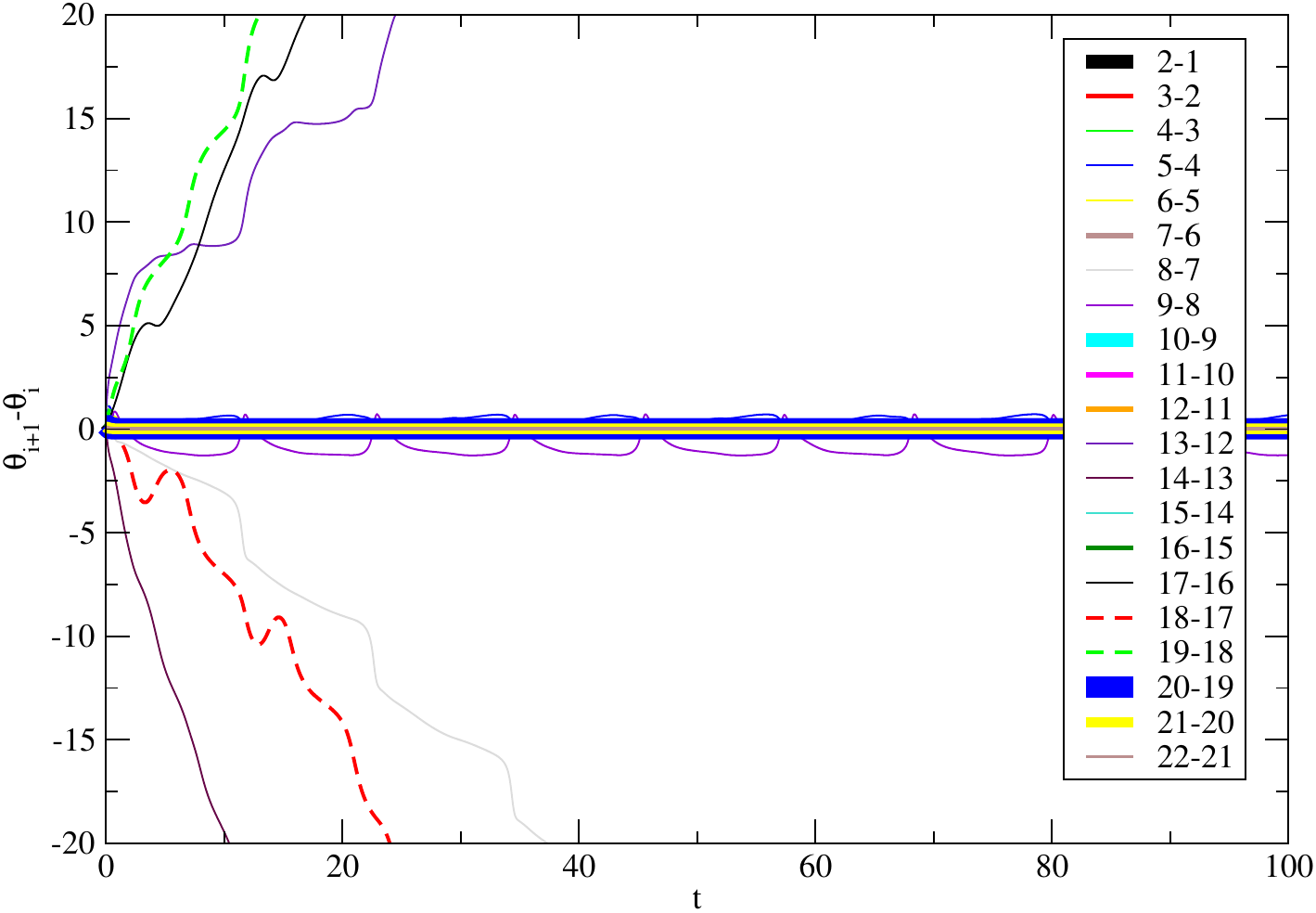}
        \includegraphics[width=0.49\columnwidth,clip]{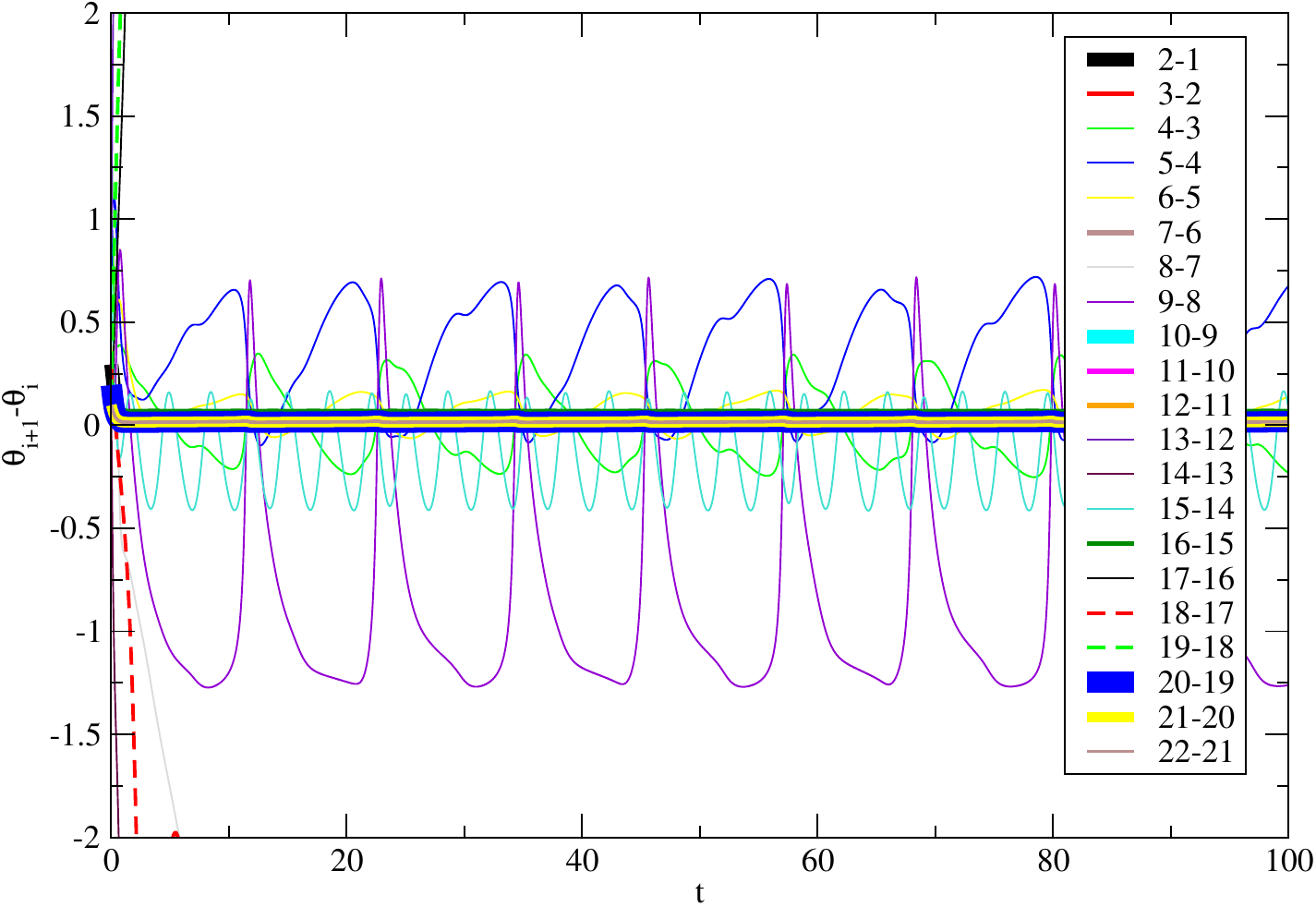}
        \includegraphics[width=0.49\columnwidth,clip]{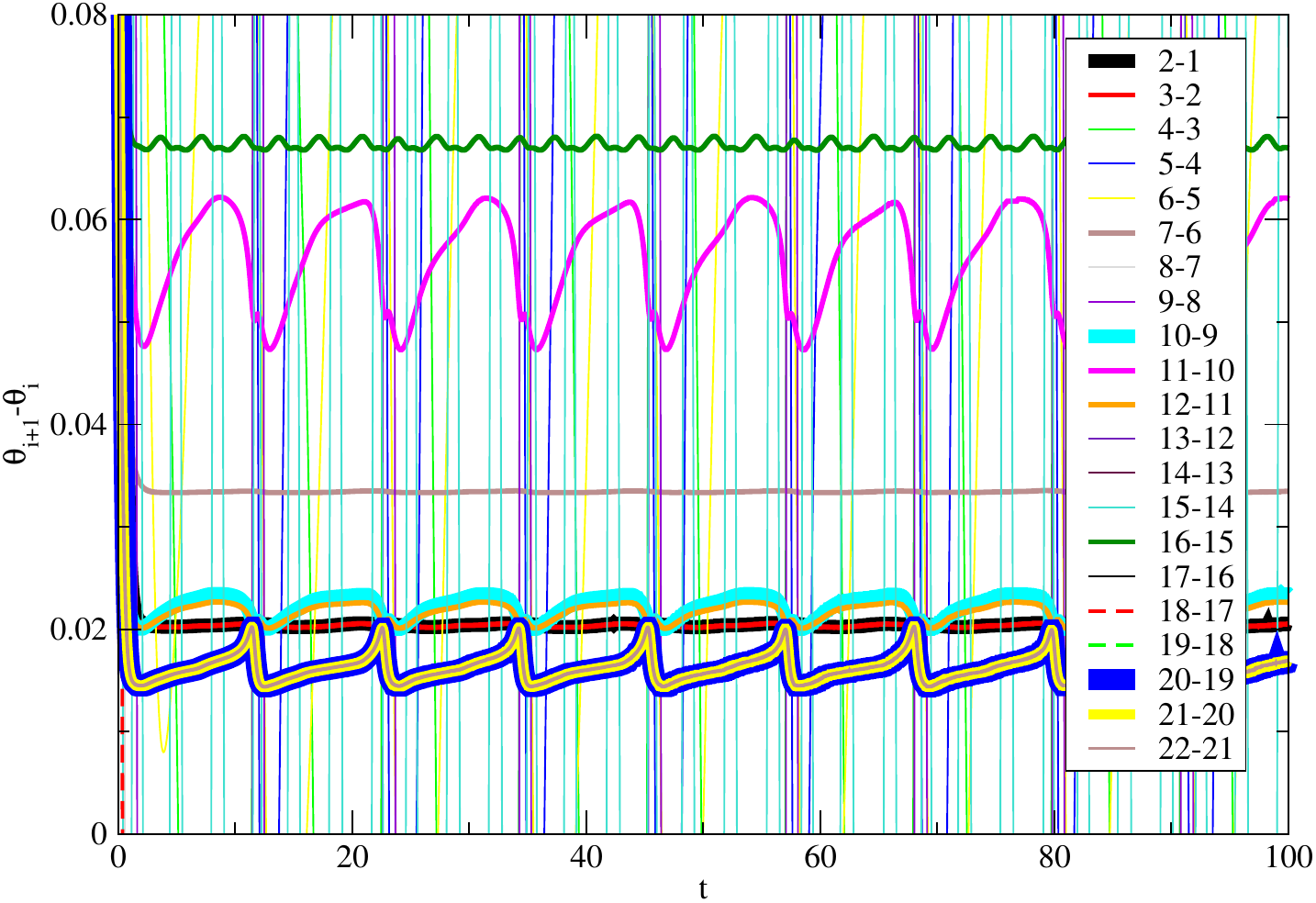}
        \caption{{Four images of Fig. 2 of MP over several scales.}}
        \label{fig2_SM}
      \end{figure}  

\begin{figure}[h]
  \centering
  \includegraphics[width=0.75\columnwidth,clip]{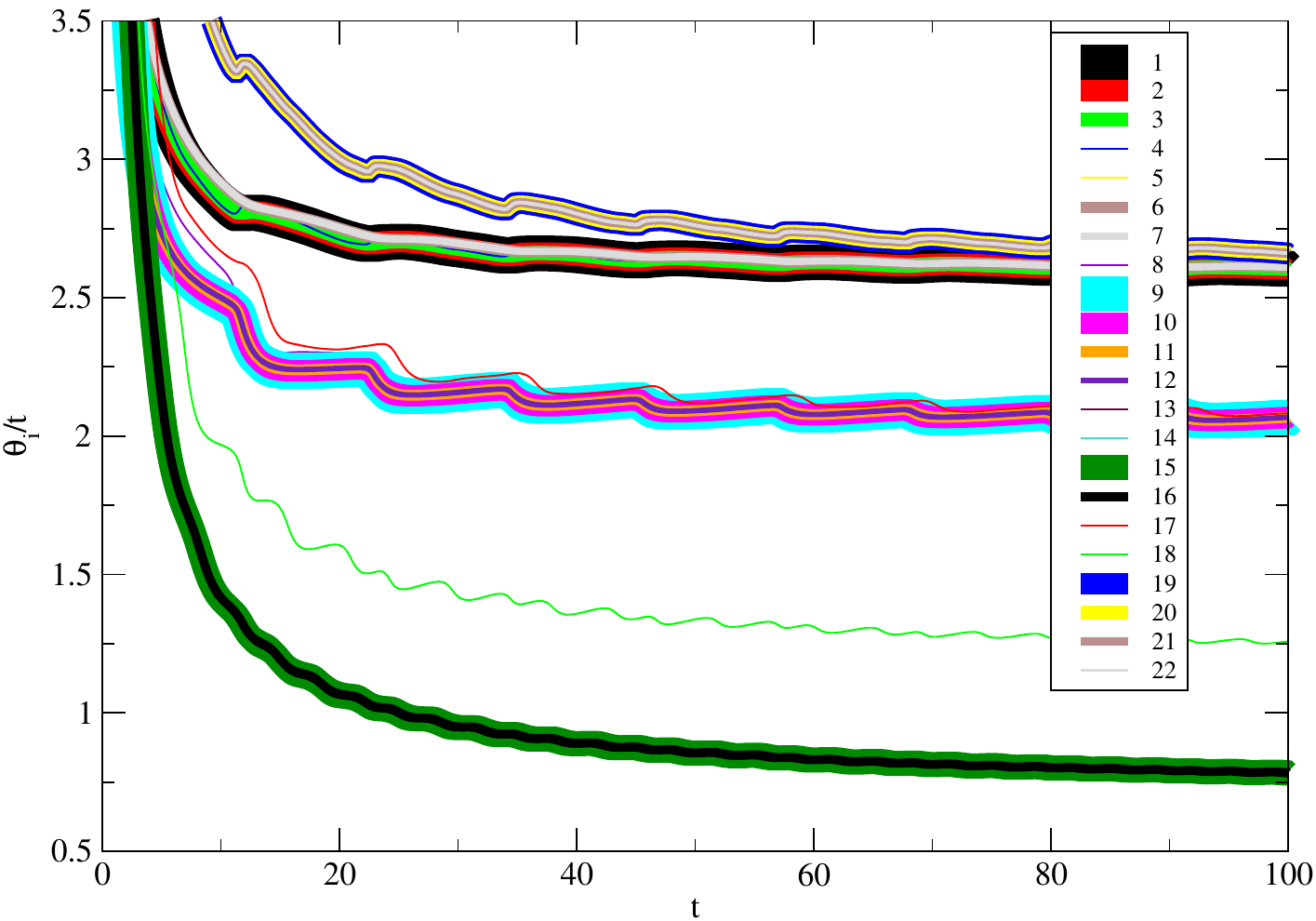}
  \caption{Rotating numbers for the same system of Fig. 2 of MP.}
  \label{fig3_SM}
\end{figure}


As for the simulations reported in Fig. 3 of MP, the initial conditions and the natural frequencies, respectively, are:\\
      $\theta_{1}=1.3;
      \theta_{2}=1.5;
      \theta_{3}=2.1;
      \theta_{4}=3.2;
      \theta_{5}=3.7;
      \theta_{6}=4.1;
      \theta_{7}=5.2;
      \theta_{8}=5.3;
      \theta_{9}=5.4;
      \theta_{10}=5.5;
      \theta_{11}=5.6;
      \theta_{12}=6.2;
      \theta_{13}=6.5;
      \theta_{14}=6.6;
      \theta_{15}=7.4;
      \theta_{16}=7.5;
      \theta_{17}=7.6;
      \theta_{18}=7.9;
      \theta_{19}=8.0;
      \theta_{20}=8.8;
      \theta_{21}=9.1;
      \theta_{22}=9.2$,
      and
      $\omega_1=1.8,\omega_2=2.4,\omega_3=3.2,\omega_4=4.5
       \omega_5=3.5,\omega_6=1.6,\omega_7=2.9,\omega_8=1.1
       \omega_9=1.2,\omega_{10}=1.8,\omega_{11}=1.3,\omega_{12}=3.0,
       \omega_{13}=2.5,\omega_{14}=0.8,\omega_{15}=1.1,\omega_{16}=1.2,
       \omega_{17}=1.5,\omega_{18}=0.8,
       \omega_{19}=3.2,\omega_{20}=3.5,\omega_{21}=2.9,\omega_{22}=2.8$.
       
      Finally, we have performed other tests confirming that the number of convergent lines in the rotating number graphs seems being related to the number of
      TE groups of the system. Specifically, by drawing 20 random samples each having the 22 frequencies $\omega_i$ drawn uniformly at random in the range $[0,5]$, with $J=1$, 
      for the number of converging lines we have obtained the mean value 4.65 with the following distribution: 2,2,3,3,4,4,4,4,4,4,5,5,5,6,6,6,6,6,7,7.
      

\end{document}